\documentclass[11pt,a4paper,fleqn]{article}

\usepackage{arxiv}
\usepackage{authblk}

\usepackage[T1]{fontenc}
\usepackage{lmodern}
\usepackage[authoryear,round]{natbib}
\usepackage{amsmath}
\usepackage{amssymb}
\usepackage{booktabs}
\usepackage{tabularx}
\usepackage{caption}
\usepackage{placeins}
\usepackage{enumitem}
\usepackage{microtype}
\usepackage{hyperref}
\hypersetup{hidelinks,hypertexnames=false,pdftitle={What Human--AI Interaction Leaves Behind: Residual Human States---A Trajectory-Based Framework and Multi-Study Analysis},pdfauthor={David Herrera-Poyatos, Carlos Pel\'aez-Gonz\'alez, Cristina Zuheros, Francisco Herrera}}

\newcommand{\X}{\mathbf{X}}
\newcommand{\Tres}{\Delta \mathbf{X}^{\mathrm{res}}}
\newcommand{\T}{\mathcal{T}}
\newcommand{\eps}{\varepsilon}

\title{What Human--AI Interaction Leaves Behind: Residual Human States---A Trajectory-Based Framework and Multi-Study Analysis}
\newcommand{\shorttitle}{Residual Human States under Human--AI Interaction}

\author{David Herrera-Poyatos$^{1*}$, Carlos Pel\'aez-Gonz\'alez$^{1}$, Cristina Zuheros$^{1}$, Francisco Herrera$^{1,2}$}

\affil{$^{1}$Department of Computer Science and Artificial Intelligence, Andalusian Institute of Data Science and Computational Intelligence (DaSCI), University of Granada, Spain. \\ Emails: \texttt{\{divadhp, carlosprog, czuheros\}@ugr.es}, \texttt{herrera@decsai.ugr.es}}

\affil{$^{2}$ADIA Lab, Abu Dhabi, United Arab Emirates \\ $^{*}$Corresponding author.}

\date{\today}

\begin{document}

\maketitle

\begin{abstract}
Human--AI Interaction is commonly evaluated by what an artificial agent enables during use, yet interaction can also alter the human who acts next. We introduce Residual Human States, a trajectory-based framework for representing the potentially multidimensional human state carried forward after interaction with AI, relative to a relevant alternative trajectory. The framework distinguishes assisted output from human-state transition and derives four empirical propositions: multidimensionality, interaction-policy dependence, horizon dissociation, and temporal persistence. Its central claim is that assisted performance does not by itself determine the human state carried forward.

We examine these propositions through four empirical cases with different inferential roles. Published medical-training evidence illustrates how one AI interaction history can be associated with both slightly higher observed accuracy on some later unaided cases and transmission of a systematic decision bias on others. A second diagnostic-learning study shows that later unaided behaviour differs across AI-support policies, while also revealing that aggregate accuracy cannot by itself distinguish improved discrimination from a shifted response criterion. In a classroom experiment, generative-AI tutoring substantially improved assisted practice performance without a corresponding detectable gain on a subsequent unaided exam. A longitudinal reanalysis of personalised AI dialogue further shows that residual contrasts can evolve across post-interaction horizons and that such change may arise from the comparison trajectory as well as from the focal intervention.

Evaluating interactive AI systems therefore requires attention not only to contemporaneous performance, but also to which human states alternative interaction trajectories produce, relative to which counterfactuals, and how durably.
\end{abstract}


\section{Introduction}
\label{sec:introduction}

Artificial intelligence (AI) is increasingly evaluated through two closely related questions: what the model can do, and what humans can accomplish with it in a Human--AI Interaction (HAI) context. Both remain essential. Yet they share a temporal boundary: evaluation commonly ends with the assisted episode. That boundary becomes incomplete when interaction with AI changes the person who will later reason, decide, learn, challenge, transfer, or intervene.

Recent evidence shows why this distinction matters. Generative-AI support can raise practice performance without a detectable corresponding gain in later unaided performance under some interaction policies \citep{bastani2025}; AI-supported practice can also produce positive transfer \citep{lira2025}; AI interaction can transmit systematic bias into later unaided decisions, alongside slightly higher observed accuracy on other cases \citep{vicente2025}; and later unaided diagnostic accuracy can differ across ways of structuring AI support, particularly whether trial-level feedback is available \citep{cabitza2026}. Other work shows that improved AI-assisted reasoning need not be accompanied by correspondingly accurate metacognitive monitoring \citep{fernandes2026}, while longitudinal intervention evidence demonstrates that differences created by AI-mediated support can change across immediate and delayed post-interaction horizons \citep{corbett2026}. The common lesson is not that AI assistance uniformly improves or erodes human capability. It is that \emph{assisted performance does not determine the human state carried forward}.

A closely related recent account is the concept of \emph{fluent dependence} proposed by \citet{anderson2026fluent}, which identifies a validity problem arising when AI-mediated output satisfies evaluative criteria without corresponding development of the human capability that the output appears to demonstrate. Their framework makes the output--capability gap explicit and proposes paired AI-mediated and AI-withdrawn tasks as a diagnostic strategy. The present paper addresses a broader trajectory question. Rather than asking only whether mediated performance overstates human capability, we ask how alternative histories of Human--AI Interaction leave differences across multiple dimensions of the human state, relative to relevant alternative trajectories, and how those differences evolve across post-interaction horizons.

Several established literatures explain parts of this divergence. Cognitive offloading shows why distributing cognitive work across external resources can be adaptive rather than pathological \citep{risko2016}; extended-cognition accounts similarly resist treating unaided performance as the only legitimate benchmark \citep{clark2025}. Research on trust and reliance shows that advice-taking changes with experience \citep{lee2004,chacon2022,eckhardt2025}, while work on learning and skill distinguishes current task success from later acquisition, transfer, or decay \citep{yan2024,macnamara2024}. More recent work makes temporality explicit through algorithmic habituation, repeated-exposure designs, path-dependent interaction, and coupled human--AI dynamics \citep{herrera2026,mladin2026,wang2026,zhu2026,park2026,zhou2026}. Each of these literatures, however, addresses a particular outcome, mechanism, or timescale; they do not yet provide a common way to describe how several dimensions of the human state differ after one interaction history compared with another.

The unresolved problem is therefore one of representation and empirical coordination, broader than either longitudinal exposure or the validity of inferring human capability from AI-mediated output: how should we describe the multidimensional human state produced by one history of human--AI interaction relative to another?

This paper proposes a trajectory-based framework for that higher-level problem. Its central object is the \emph{residual human state}: the role-relevant, potentially multidimensional state carried forward by the human after a history of HAI. Rather than asking only what a human--AI configuration produces at episode $t$, we distinguish the observable output $Y_t$ from the human state $\X_{t+1}$ that enters the next episode.

The paper makes three contributions through the Residual Human States framework. First, it introduces a common state-transition representation in which competence, judgment, reliance, motivation or metacognition, bias, agency, and other domain-relevant dimensions can change simultaneously without being assumed commensurable. Second, it extends the evaluation object beyond the output--capability relation by defining residual human-state change relative to an alternative interaction trajectory and by making post-interaction horizon explicit, from which four empirical propositions are derived concerning multidimensionality, policy dependence, horizon dissociation, and temporal persistence. Third, it coordinates four empirical cases within this trajectory-relative framework while deliberately matching analytical depth to the inferential role of each case. The four cases are not treated as estimates of a common effect size or as equally strong tests of a common latent construct.

The contribution is therefore not the generic claim that AI use can affect later human behaviour, nor that such effects may change over time. It is the coordination of these phenomena within a trajectory-relative representation that distinguishes assisted output from the human state carried forward, makes the comparison trajectory part of the estimand, and indexes residual differences explicitly by post-interaction horizon.

The guiding question is:
\begin{quote}
\emph{Which interaction trajectories produce which human states, relative to which alternatives, and with what persistence or recovery?}
\end{quote}

Section~\ref{sec:stateoftheart} establishes the representational gap. Section~\ref{sec:framework} develops the trajectory-based residual human-state method and derives four empirical propositions. Section~\ref{sec:empirical-framework} specifies the coordinated secondary-analysis design. Section~\ref{sec:results} reports the proposition-specific results and cross-study synthesis. Section~\ref{sec:discussion} discusses implications for HAI evaluation, interaction policy, agency, and governable dependence. Section~\ref{sec:conclusion} concludes.

\section{State of the Art}
\label{sec:stateoftheart}

The purpose of this section is not to provide a comprehensive review of Human--AI Interaction, but to identify the conceptual elements required to represent what interaction leaves in the human after assistance. Several established literatures already explain important parts of this problem. Cognitive offloading and extended cognition clarify why dependence on external resources is not equivalent to deterioration; research on trust and reliance shows that interaction changes how humans use algorithmic advice; learning, bias, and metacognition demonstrate that different human outcomes can diverge; and longitudinal work on habituation, repeated exposure, and path dependence establishes that these effects unfold over time.

Taken together, however, these literatures do not yet provide a common representation of the human state produced by an interaction history. They typically focus on particular outcomes, mechanisms, or temporal processes rather than on how multiple human dimensions can be compared across alternative interaction trajectories and post-interaction horizons. The subsections that follow therefore review the literature selectively around this representational problem. Each subsection is organised around one guiding question: how to distinguish a changed human from a missing external resource (Section~\ref{subsec:offloading}); whether different dimensions of the human state change together or diverge, and whether this depends on task and domain (Section~\ref{subsec:adaptation}); and how such changes depend on interaction history and evolve after interaction ends (Section~\ref{subsec:temporal}). The final subsection identifies the representational gap that motivates the Residual Human States framework.

\subsection{Cognitive offloading, extended cognition, and assisted performance}
\label{subsec:offloading}

\emph{When performance falls after AI is removed, how can we tell whether the human has changed or whether a useful external resource is simply missing?} Answering this question requires distinguishing functional dependence on an external cognitive resource from a change in the human produced by the history of using that resource. This distinction is essential for residual human-state analysis because a decline in performance when assistance is removed may reflect the loss of a useful external component rather than deterioration of the human state itself.

The idea that cognition can be reorganised through technological coupling long predates contemporary AI. Licklider's man--computer symbiosis framed humans and computers as contributors to a joint process \citep{licklider1960}; the extended-mind thesis later argued that external resources can, under appropriate conditions, become constitutive of cognitive activity \citep{clarkchalmers1998}. Generative AI makes this perspective newly salient because external systems now participate in formulation, explanation, synthesis, planning, and reasoning \citep{clark2025}.

Cognitive offloading provides the direct psychological foundation. External resources can alter task demands, reduce unnecessary burden, and improve performance \citep{risko2016}. This establishes an important boundary for residual human-state analysis: poorer performance after a useful resource is removed is not, by itself, evidence that prior use damaged human competence. Functional dependence on a tool must be distinguished from a change in the human produced by the history of using it.

Evaluation has often approached this coupling through contemporaneous outcomes: whether human--AI combinations outperform humans alone, AI alone, or the stronger component. Meta-analytic evidence shows substantial heterogeneity and cautions against assuming synergy \citep{vaccaro2024}. Its primary object, however, remains configuration performance while human and AI operate together, not the capability, strategy, disposition, or metacognitive state carried by the human into later activity.

\subsection{Learning, reliance, bias, metacognition, and human adaptation}
\label{subsec:adaptation}

\emph{Do different dimensions of the human state change together after AI interaction, or can they diverge, and does this depend on the task and domain in which the interaction takes place?} Once functional dependence is distinguished from human-state change, a second problem appears: interaction with AI can alter several human outcomes that need not move together. Competence, reliance, bias, metacognition, and related dimensions therefore need to be considered separately before they can be coordinated within a common post-interaction representation.

Trust and reliance research has long shown that reliance should be calibrated to system capability and context rather than maximised indiscriminately \citep{lee2004}. More recent HAI work treats reliance as dynamic and sensitive to advice quality, task structure, and interaction design \citep{eckhardt2025}. Longitudinal evidence further shows that advice-taking changes with repeated exposure \citep{chacon2022}.

The same temporal distinction appears in learning and skill. \citet{bastani2025} report that unrestricted generative-AI assistance can substantially raise practice performance while reducing later unaided mathematics performance, whereas a scaffolded tutor mitigates much of that effect; as revisited in Section~\ref{subsec:results-p3}, the reduction under unrestricted assistance is sensitive to small-sample cluster-robust inference, whereas for the scaffolded tutor no later difference was detected despite large assisted gains. \citet{lira2025} report improved later unaided writing after AI-supported practice. In medical training, \citet{vicente2025} show that trainees who received AI advice reproduced an AI system's systematic bias in later unaided decisions, together with slightly higher observed accuracy on cases where the AI had been correct. \citet{cabitza2026} further report a policy-specific difference in later
unaided accuracy on cases governed by a hidden diagnostic criterion, which
the authors interpret as evidence of knowledge transfer. As discussed in Section~\ref{subsec:results-p2},
the observed accuracy probe also admits an alternative response-criterion
interpretation, so the present paper uses this study primarily to test
interaction-policy dependence rather than to identify the underlying
learning mechanism.

Metacognition adds another dimension. \citet{fernandes2026} report that LLM assistance improves logical-reasoning performance while users remain poorly calibrated in judging their own performance, demonstrating that better assisted output and better self-monitoring need not move together.

\citet{anderson2026fluent} integrate one important subset of these findings under the construct of \emph{fluent dependence}, which reframes AI-mediated performance as a construct-validity problem (Section~\ref{sec:introduction}). Residual human-state analysis has a different object: the difference in human state produced by one interaction trajectory relative to another, which may involve competence but also judgment, reliance, metacognition, internalised bias, and agency, and does not require the comparison trajectory to be AI-free.

Together, these findings show that competence, reliance, bias, and metacognition can respond differently to AI-mediated interaction. A general account therefore cannot use any one of these variables as a universal proxy for the human state. The same evidence also spans mathematics, writing, medical diagnosis, and logical reasoning, which raises the further question of whether such divergence depends on task and domain. Because these studies also differ in interaction policy, population, outcome, and design, they cannot isolate a domain effect. Within the framework developed below, task and domain enter through the contextual conditions $K_t$ (Section~\ref{subsec:state-transition}), so domain dependence is treated as an open empirical question rather than as an established property.

\subsection{Repeated exposure, path dependence, persistence, and recovery}
\label{subsec:temporal}

\emph{How do these changes depend on the history of interaction, and do they persist, fade, or recover relative to an alternative trajectory once interaction ends?} A multidimensional account also requires an explicit temporal structure. Residual differences can depend on the order, repetition, and duration of interaction, and they may persist, attenuate, strengthen, converge, or recover after the AI-containing episode has ended. The relevant problem is therefore not only whether interaction history matters, but how trajectory-relative differences evolve across post-interaction horizons.

Habituation provides one explanatory vocabulary for repeated interaction. Passive reliance can become normalised through repeated HAI, with potential consequences for deliberation, judgment, autonomy, intention, and second-order agency \citep{herrera2026}. Algorithmic habituation has likewise been framed as gradual accommodation to recurrent AI-mediated patterns \citep{mladin2026}. Repeated-exposure methodology and large-scale user--LLM evidence further show why exposure histories and early interaction patterns matter \citep{wang2026,zhu2026}.

Path dependence is related but broader than habituation. Formal models now represent human capability and delegation as coupled variables and analyse longer-term human--AI coevolution across multiple timescales \citep{park2026,zhou2026}. These developments establish that neither ``time matters'' nor ``interaction history matters'' is a novel proposition by itself.

What remains less systematically integrated is the evolution of a trajectory contrast after interaction. \citet{corbett2026} provide an especially useful example: a personalised corrective AI dialogue produced a larger immediate reduction in psychology misconceptions than a textbook refutation or neutral AI dialogue, and both corrective interventions remained superior to the neutral control two months later. The published longitudinal analysis already tested a Condition $\times$ Time interaction on a common complete-case sample; however, the comparisons highlighted in the article combine samples and adjustments, and comparator-specific changes in the contrast were not estimated with their uncertainty. This pattern illustrates why persistence, decay, and convergence should be represented as empirical properties of a residual contrast rather than assumed from an immediate post-test.

\subsection{The representational gap}
\label{subsec:gap}

The framework proposed here also builds on established traditions rather than replacing them. The distinction between performance during acquisition and what is retained afterwards is central to research on learning versus performance and desirable difficulties \citep{soderstrom2015}; the horizon-dissociation proposition developed in Section~\ref{subsec:propositions} transposes it to settings in which the assisted output is produced jointly with an AI system. Work on the ironies of automation and on automation complacency and bias showed that automated support can let the manual and cognitive skills needed when automation fails deteriorate through disuse \citep{bainbridge1983}, and can reduce operators' monitoring of automated aids and bias their decisions towards automated recommendations \citep{parasuraman2010}, and recent observational evidence from colonoscopy reports lower unaided adenoma detection after routine exposure to AI assistance \citep{budzyn2025}, a pattern of the kind the framework is designed to represent. Formally, the residual contrast defined in Section~\ref{sec:framework} is a contrast between potential outcomes under two treatment trajectories evaluated at a given horizon, in the sense of causal inference for time-varying treatments \citep{robins1986,hernan2020}. What the present framework adds is not this causal notation, but the coordination of several human-state dimensions in one object, the explicit treatment of the comparison trajectory as part of the estimand, and the decomposition of temporal change into focal and comparator movement.

The literature therefore leaves us beyond a generic ``longitudinal gap.'' HAI is temporally adaptive, assisted performance can diverge from later human outcomes, interaction policy can matter, and residual change is not uniformly positive or negative. What remains missing is a higher-level representational layer connecting these findings. Three requirements recur:

\begin{enumerate}[leftmargin=1.6em,label=\textbf{G\arabic*.}]
\item \textbf{A common human-state object.} The affected human cannot be represented adequately by performance, reliance, or competence alone when different dimensions may change simultaneously and in different directions.
\item \textbf{A trajectory-relative contrast.} Post-interaction change cannot generally be inferred from a simple pre--post comparison because the relevant comparison trajectory also evolves through practice, time, and experience. The counterfactual must therefore be another trajectory rather than an immutable baseline.
\item \textbf{A cross-domain temporal language.} The framework must accommodate learning, habituation, bias, metacognition, skill change, persistence, and recovery without asserting a single mechanism or universal timescale.
\end{enumerate}

Together, these requirements motivate a trajectory-based representational layer in which the human state is treated as a distinct evaluation object, residual change is defined relative to an alternative interaction trajectory, and post-interaction horizon is made explicit.

\section{Method: A Trajectory-Based Residual Human-State Framework}
\label{sec:framework}

The purpose of the framework is to provide a common analytical language for human changes that are currently studied through separate constructs such as learning, reliance, bias, metacognition, habituation, and recovery. Rather than treating these phenomena as manifestations of a single psychological mechanism, we represent them as potentially distinct changes in a multidimensional human state produced by an ordered history of Human--AI Interaction. The method therefore proceeds in four steps. First, it separates assisted task performance from the human state carried forward after interaction. Second, it represents that state as the result of an interaction trajectory rather than a single episode. Third, it defines residual human-state change relative to a relevant alternative trajectory and makes post-interaction time explicit. Finally, it derives observable residual-state regimes and four empirical propositions that can be instantiated using domain-specific measures. The resulting framework is thus a representational and measurement-oriented method: it specifies the objects to be compared and the contrasts to be estimated without assuming that different human-state dimensions are directly observable, commensurable, or generated by the same underlying process.

\subsection{Human state, assisted performance, and state transition}
\label{subsec:state-transition}

We treat HAI as a sequence of state-changing encounters. Let
$t=0,1,\ldots,T-1$ index the ordered interaction episodes, and let
$\X_t\in\mathcal{X}$ denote the role-relevant human state immediately
before episode $t$. The interaction sequence therefore generates the
state trajectory $\X_0,\X_1,\ldots,\X_T$, where $\X_T$ is the state
carried forward after the final interaction episode.

An illustrative representation is
\begin{equation}
\X_t=(C_t,J_t,R_t,M_t,B_t,A_t),
\label{eq:state}
\end{equation}
where $C_t$ may represent autonomous competence, $J_t$ judgment or problem-solving strategy, $R_t$ reliance propensity, $M_t$ motivational or metacognitive state, $B_t$ internalised biases or learned dispositions, and $A_t$ dimensions of agency or practical control. 
The vector is not proposed as a universal psychological ontology or as a latent-variable model intended to estimate the complete state $\X_t$. Rather, it specifies a representational space of role-relevant human-state components that an evaluation may need to distinguish; each empirical domain can instantiate only the subset and measurement model that are substantively justified and for which defensible observable probes are available.

Let $Y_t$ denote the observable task outcome generated during episode $t$, $Z_t$ the AI system or system state, $P_t$ the interaction policy through which assistance is delivered, and $K_t$ the task and contextual conditions. Current assisted performance can be represented schematically as
\begin{equation}
Y_t=\mathcal H(\X_t,Z_t,P_t,K_t).
\label{eq:performance}
\end{equation}

The interaction may also change the human state entering the next episode:
\begin{equation}
\X_{t+1}=\T_{Z_t,P_t,K_t,\Delta t_t}(\X_t),
\label{eq:transition}
\end{equation}
where $\T$ is a human-state transition operator and $\Delta t_t$ denotes elapsed time where relevant. Equation~\eqref{eq:transition} is an analytical representation, not a claim that the transition is linear, stationary, deterministic, or directly observable.

The framework begins with a separation:

\[
\text{current output}\neq\text{human state produced for later activity}.
\]

An interaction can improve $Y_t$ without increasing $C_{t+1}$, increase competence while also increasing an undesirable bias component, or leave competence approximately unchanged while modifying reliance or metacognitive calibration.

\subsection{Interaction trajectories and residual human-state effects}
\label{subsec:residual-effects}

An interaction trajectory is an ordered sequence
\begin{equation}
\Gamma_{0:T-1}=\big[(Z_0,P_0,K_0,\Delta t_0),\ldots,(Z_{T-1},P_{T-1},K_{T-1},\Delta t_{T-1})\big].
\label{eq:trajectory}
\end{equation}

For brevity, we write $\Gamma \equiv \Gamma_{0:T-1}$ when the interaction
interval is clear from context. The state reached after that trajectory is
\begin{equation}
\X_T=\Phi(\X_0,\Gamma).
\label{eq:evolution}
\end{equation}

A state reached after AI interaction is not informative by itself because humans also change without AI through practice, fatigue, forgetting, feedback, and exposure to other resources. Let $\Gamma$ denote an interaction trajectory of interest and $\Gamma^c$ a relevant comparison trajectory. We define the \emph{residual human-state effect} as
\begin{equation}
\Tres(\Gamma,\Gamma^c)=\Phi(\X_0,\Gamma)-\Phi(\X_0,\Gamma^c)
\label{eq:residual}
\end{equation}

Where state components are numerical, subtraction is componentwise. Where a component is not naturally numeric, Equation~\eqref{eq:residual} is shorthand for an explicitly defined domain-specific contrast rather than a universal metric over $\mathcal X$. Equation~\eqref{eq:residual} defines the trajectory-relative contrast at the conceptual individual level. Individual counterfactual states under $\Gamma$ and $\Gamma^c$ are not jointly observed; in between-group empirical designs the corresponding estimand is an average contrast over the distribution of baseline states, $\mathbb{E}_{\X_0}\!\left[\Phi(\X_0,\Gamma)-\Phi(\X_0,\Gamma^c)\right]$, which can also be understood as averaging over any stochastic variation in the state transitions and in the realised trajectory, probed through the observable outcomes described in Section~\ref{sec:empirical-framework}.

The comparison trajectory need not be ``no AI.'' It may represent a different interaction policy, delayed assistance, human-first deliberation, alternative explanatory support, or another practically meaningful configuration. This makes $P_t$ part of the intervention/trajectory object rather than a peripheral interface detail.

\subsection{Path dependence, persistence, and recovery}
\label{subsec:path-persistence}

A testable form of path dependence occurs when the order of otherwise comparable interaction conditions changes the resulting state. For two transition types $A$ and $B$,
\begin{equation}
\T_B\circ\T_A\neq\T_A\circ\T_B.
\label{eq:path}
\end{equation}

This non-commutativity is one empirically testable form of trajectory dependence; it is not a definition of all possible path dependence.

Residual effects may also evolve after the AI-containing portion of a trajectory ends. 
Let $\X_{\tau}(\Gamma)$ denote the human state at post-interaction
horizon $\tau$ after trajectory $\Gamma$. We therefore make the post-interaction horizon explicit:
\begin{equation}
\Tres(\Gamma,\Gamma^c;\tau) =
\X_{\tau}(\Gamma)-\X_{\tau}(\Gamma^c),
\label{eq:tau}
\end{equation}
where $\tau$ indexes the elapsed post-interaction horizon at which the relevant state probe is observed. In $\X_{\tau}(\Gamma)$ the subscript denotes elapsed post-interaction time rather than an episode index; immediately after the final episode, $\X_{\tau}(\Gamma)$ coincides with the state $\Phi(\X_0,\Gamma)$ reached in Equation~\eqref{eq:evolution}, and it may continue to change afterwards.

Different components can persist, decay, strengthen, or recover at different rates. Because the comparison trajectory also continues to evolve after interaction, a change in $\Tres(\Gamma,\Gamma^c;\tau)$ across horizons can originate in either trajectory:
\begin{equation}
\Tres(\Gamma,\Gamma^c;\tau_2)-\Tres(\Gamma,\Gamma^c;\tau_1) =\big[\X_{\tau_2}(\Gamma)-\X_{\tau_1}(\Gamma)\big]-\big[\X_{\tau_2}(\Gamma^c)-\X_{\tau_1}(\Gamma^c)\big].
\label{eq:tau-decomposition}
\end{equation}

A shrinking residual contrast is an attenuation of the effect relative to that comparator, but it is not by itself evidence that the outcome within the focal trajectory deteriorated by the same amount.

\subsection{Residual-state signatures and regimes}
\label{subsec:signatures}

To compare multidimensional effects without assuming that state components share a common scale, define
\begin{equation}
s_{\eps_j}(x)=
\begin{cases}
+,&x>\eps_j,\\
0,&|x|\leq\eps_j,\\
-,&x<-\eps_j,
\end{cases}
\label{eq:sign}
\end{equation}
where $\eps_j$ is a component-specific region of practical equivalence; empirically, assigning the value $0$ requires an equivalence test against $\pm\eps_j$ rather than a non-significant difference; by the same logic, assigning $+$ or $-$ should take the uncertainty of the estimate into account relative to $\pm\eps_j$, rather than rely on the point estimate alone. The residual-state signature is
\begin{equation}
\Sigma_{\boldsymbol{\eps}}(\Gamma,\Gamma^c)=\left(s_{\eps_1}(\Delta X^{\mathrm{res}}_1),\ldots,s_{\eps_m}(\Delta X^{\mathrm{res}}_m)\right).
\label{eq:signature}
\end{equation}

The signature is descriptive rather than normative. A positive change in competence and a positive change in an undesirable bias variable have different meanings.

Let $\mathcal A$ denote assisted episodes over which current task performance is evaluated and $\bar Y_{\mathcal A}(\Gamma)$ a matched summary of performance. Define
\begin{equation}
G(\Gamma,\Gamma^c)=\mathbb E[\bar Y_{\mathcal A}(\Gamma)]-\mathbb E[\bar Y_{\mathcal A}(\Gamma^c)].
\label{eq:G}
\end{equation}

The two-horizon representation is
\begin{equation}
\mathcal E(\Gamma,\Gamma^c)=\left(G(\Gamma,\Gamma^c),\Sigma_{\boldsymbol{\eps}}(\Gamma,\Gamma^c)\right).
\label{eq:twohorizon}
\end{equation}

Equation~\eqref{eq:twohorizon} is a representational device rather than an identification strategy.

Four recurrent empirical regimes can then be represented. \emph{Appropriation} occurs when a role-relevant competence is greater under the AI-mediated trajectory than under the comparison trajectory. \emph{Preservation} describes improved assisted performance with a later competence difference that lies within a justified equivalence margin; improved assisted performance with a later difference that is merely not detected, without such a test, is reported as \emph{undetected later difference} rather than as preservation. \emph{Displacement} describes a lower later competence probe under the AI-containing trajectory. \emph{Multidimensional transformation} occurs when different state components change simultaneously and potentially in different evaluative directions. These regime labels describe trajectory-relative patterns rather than universal psychological mechanisms. In particular, preservation and displacement are always defined relative to the specified comparison trajectory: a residual difference near zero is compatible with both trajectories changing by the same amount, and does not imply the absence of absolute change in either.

\subsection{Empirical propositions derived from the framework}
\label{subsec:propositions}

The framework yields four propositions that organise the empirical analysis.

\paragraph{P1: Multidimensionality.}
Alternative interaction trajectories may produce simultaneous residual consequences that differ in substantive and evaluative meaning. The representational implication is that post-interaction change need not be adequately described by a single scalar outcome: one trajectory may improve one observed dimension while worsening another. Whether such multidirectional consequences reflect distinct latent components of the human state is a separate empirical question. A common underlying mechanism may generate different observed consequences across tasks or case classes, and distinct state components may themselves be correlated. P1 is therefore a representational proposition about the possibility and evaluative relevance of multidimensional residual change; the empirical case examined below establishes the need for a non-scalar description of observed consequences, not the independence of multiple latent human-state components.

\paragraph{P2: Policy dependence.} \emph{Residual outcomes can depend on how AI capability is delivered, not merely on whether AI is available.} Formally, for policies $P_1\neq P_2$, the induced trajectories can satisfy
\[
\Tres(\Gamma(P_1),\Gamma^c)\neq\Tres(\Gamma(P_2),\Gamma^c).
\]

\paragraph{P3: Horizon dissociation.} \emph{Current assisted performance does not by itself identify subsequent human capability.} In particular,
\begin{equation}
G(\Gamma,\Gamma^c)>0\;\nRightarrow\;\Delta C^{\mathrm{res}}(\Gamma,\Gamma^c)>0.
\label{eq:horizon-proposition} 
\end{equation}

Here $\Delta C^{\mathrm{res}}$ denotes the competence component $C$ of $\Delta\X^{\mathrm{res}}$ in Equation~\eqref{eq:residual}. Equation~\eqref{eq:horizon-proposition} states a structural possibility of the framework: assisted performance and later capability are distinct quantities, so the first does not fix the second. Empirically, the relevant claim is weaker and epistemic. A study such as Study~3 does not establish a causal counterexample to the implication; it can show that a large assisted gain is insufficient to infer the sign or magnitude of $\Delta C^{\mathrm{res}}$.

\paragraph{P4: Temporal persistence.} \emph{Residual trajectory contrasts may persist, decay, strengthen, or converge across post-interaction horizons.} Thus, for at least some trajectories and state components,
\begin{equation}
\Tres(\Gamma,\Gamma^c;\tau_1)\neq\Tres(\Gamma,\Gamma^c;\tau_2).
\label{eq:temporal-proposition}
\end{equation}

A mediated--unmediated performance gap, of the type captured by fluent dependence \citep{anderson2026fluent}, is conceptually distinct from a trajectory-relative residual effect. Let
\[
D^{FD}=Y^{AI}-O_C^{\mathrm{withdrawn}}
\]
denote an output--capability gap measured through paired tasks. By contrast,
\[
\Delta O_C^{\mathrm{res}}(\Gamma,\Gamma^c) = \mathbb{E}[O_C(\Gamma)-O_C(\Gamma^c)]
\]
compares the later observable competence probe produced by two alternative interaction histories. A large $D^{FD}$ does not by itself identify $\Delta O_C^{\mathrm{res}}$: the former is a validity signal concerning what mediated output evidences about current human capability, whereas the latter is a trajectory effect concerning what alternative interaction histories leave in the human.

\section{Empirical Application of the Residual Human States Framework: Four Studies}
\label{sec:empirical-framework}

This section translates the representation into an empirical strategy. For each of four cases selected to examine complementary implications of Propositions P1--P4, we identify the interaction trajectory, the comparison trajectory, the observable post-interaction probe, and the horizon at which that probe is observed, preserving the domain-specific meaning of the original outcomes.

The cases deliberately play different inferential roles. Study~1 (P1) uses published
post-interaction aggregate evidence to illustrate why a scalar description
of residual change can be inadequate. Study~2 (P2) uses released participant-level data
to estimate direct between-policy contrasts on a subsequent unaided behavioural
probe. Study~3 (P3) reproduces and extends the original clustered regression analysis
with direct policy contrasts and small-sample cluster-robust inference. Study~4 (P4)
reconstructs the released longitudinal data and estimates comparator-specific
temporal contrasts under a common repeated-measures specification. The
empirical objective is therefore structured triangulation rather than
methodological symmetry: analytical depth is matched to the proposition and
to what each original design can identify.

\subsection{Study selection and common analytical strategy}
\label{subsec:study-selection}

The empirical component is a theory-driven coordinated secondary analysis, not a systematic review or meta-analysis. Studies are selected because they satisfy four criteria: (i) an intervention contrast involving AI-mediated interaction; (ii) an observable human outcome during or after interaction that can be mapped to a domain-specific state probe; (iii) publicly accessible data or analysis materials; and (iv) complementary alignment with one of Propositions P1--P4. The studies differ in domain, outcome scale, intervention, and temporal horizon; no pooled effect size is assumed or estimated. The purposive selection is intended to demonstrate how the framework can be operationalised across distinct designs; it does not estimate how frequently these patterns occur in the wider HAI literature. The analysis code was developed with the assistance of a generative AI tool (Claude, Anthropic); all scripts were reviewed by the authors, and the secondary analyses of Studies~2--4 are reproduced from the released data by the code described in the Data and code availability section.

Table~\ref{tab:theory-data-mapping} makes the translation from the theoretical framework to the empirical analyses explicit. For each proposition, it identifies the focal interaction trajectory, the relevant comparison trajectory, the human-state component represented, the observable study-specific probe, and the post-interaction horizon at which that probe is assessed. The mapping is not intended to imply that the outcomes are metrically comparable across studies. Its purpose is to show how the same trajectory-relative logic is instantiated through different experimental designs.

\begin{table*}[ht]
\centering
\scriptsize
\renewcommand{\arraystretch}{1.12}
\caption{Theory-to-data mapping for the four core empirical analyses. Each study operationalises one proposition using domain-specific interaction trajectories, comparison conditions, human-state components, observable probes, and post-interaction horizons. The mappings do not imply metric comparability across studies.}
\label{tab:theory-data-mapping}

\begin{tabularx}{\textwidth}{
>{\raggedright\arraybackslash}p{0.7cm} >{\raggedright\arraybackslash}p{1.8cm} >{\raggedright\arraybackslash}p{2.3cm} >{\raggedright\arraybackslash}p{2.2cm} >{\raggedright\arraybackslash}p{1.85cm} >{\raggedright\arraybackslash}p{2.85cm} >{\raggedright\arraybackslash}X}
\toprule
\textbf{Prop.} & \textbf{Study} & \textbf{Focal trajectory $\Gamma$} & \textbf{Comparison trajectory $\Gamma^{c}$} & \textbf{Target dimension / framework quantity} & \textbf{Observable probe} & \textbf{Post-interaction horizon} \\
\midrule

P1 & \citet{vicente2025} & AI-supported diagnostic training containing both accurate guidance and a systematic error pattern & Matched diagnostic training without AI support & Accuracy; bias$^{a}$ & Published unaided accuracy pattern across case types and reproduction /
generalisation of the AI-induced systematic response bias & Unaided post-AI phase \\

\addlinespace

P2 & \citet{cabitza2026} & AI support with trial-by-trial corrective feedback & AI advice alone and AI advice with confidence information & Response state$^{b}$ & Subsequent unaided accuracy on hidden-criterion cases & Unaided transfer phase \\

\addlinespace

P3 & \citet{bastani2025} & GPT Base or GPT Tutor during mathematics practice & Control trajectory; additionally GPT Tutor versus GPT Base for direct policy comparison & $G,\;C$ & Normalised assisted-practice performance (\texttt{Part2Tot}) and subsequent unaided-exam performance (\texttt{Part3Tot}) & Assisted practice $\rightarrow$ subsequent unaided exam \\

\addlinespace

P4 & \citet{corbett2026} & Personalised misconception-corrective AI dialogue & Textbook refutation and neutral AI dialogue & $B^{\mathrm{mis}}$\,\textsuperscript{c} & Strength of the targeted misconception / corresponding belief-correction measure & Immediate post-test, 10 days, and 2 months \\

\bottomrule
\end{tabularx}

\vspace{2pt}
{\footnotesize\raggedright$^{a}$ Candidate accuracy- and bias-related residual dimensions, not assumed to be independent latent components. $^{b}$ Domain-specific decision or response state; the probe covers cases whose correct answer is always ``positive'', so a difference may reflect a shifted response criterion. $^{c}$ Misconception-related or other domain-specific belief state.\par}

\end{table*}

The theoretical state $\X$ is latent. Each dataset provides observable probes of selected components rather than direct observation of the complete residual human-state vector. For component $j$, we write
\begin{equation}
O^j_{\mathrm{post}}=\mathcal M_j(X^j_{\mathrm{post}},K_{\mathrm{post}})+\zeta_j,
\label{eq:measurement}
\end{equation}
where $O^j_{\mathrm{post}}$ is the observed outcome, $\mathcal M_j$ the study-specific measurement relation, $K_{\mathrm{post}}$ the relevant testing context, and $\zeta_j$ measurement noise or unmodelled influence. Accordingly, empirical contrasts are reported as effects on observed probes and interpreted as evidence about selected residual-state dimensions only when the study design and measurement relation justify that interpretation.

The four empirical cases follow a common interpretive logic, while the
extent of re-estimation depends on the inferential role of each case. We:

\begin{enumerate}
    \item identify the published experimental design, comparison trajectory,
    and empirical benchmark relevant to the corresponding proposition;

    \item map the relevant outcome to an explicitly defined observable
    post-interaction probe without equating that probe with the complete
    latent human state;

    \item distinguish published illustrations from framework-specific
    secondary estimates and from deeper reanalyses of released microdata;

    \item where new contrasts are estimated, preserve the original unit of
    assignment, outcome coding, clustering structure, and substantively
    relevant covariates wherever applicable;

    \item report estimates with uncertainty and avoid inferring equivalence
    from a non-significant difference;

    \item interpret restricted behavioural probes conservatively when the
    observed outcome does not uniquely identify the underlying psychological
    mechanism; and

    \item avoid pooling across studies whose outcomes are not metrically
    commensurable.
\end{enumerate}

\subsection{Study 1: Published illustration of multidirectional residual consequences}
\label{subsec:study-vicente}

\citet{vicente2025} recruited 86 medical students and assigned them to an
AI-assisted group or an unassisted control group. During the assisted phase,
the simulated AI was accurate on most cases but systematically wrong on one
specific class of difficult cases. Both groups subsequently completed an
unaided phase. The design therefore provides a useful illustration of
post-interaction consequences because the critical outcomes are observed
after AI support has been removed.

For P1, we use the published Phase~2 aggregate results as an empirical
illustration rather than as a framework-specific participant-level
reanalysis. Two observed consequences are distinguished. The first is an
accuracy-oriented outcome on case classes for which the AI had supplied
correct guidance. The second is a bias-oriented outcome reflecting later
reproduction and generalisation of the AI's systematic error pattern. The
purpose of the case is not to establish that these outcomes correspond to
independent latent components of the human state. It is to show why one
interaction history may require more than a single scalar description when
its later behavioural consequences differ in evaluative direction.

This qualification is particularly important because the AI's systematic
error can be represented as a shifted decision threshold on the underlying
dark/light-cell ratio. A transmitted criterion could therefore improve
classification on some case classes while worsening it on others. In
addition, the published proportion tests treat multiple classification
responses as independent although responses are nested within participants.
We consequently use the published estimates descriptively as an illustration
of multidirectional residual consequences and do not rely on their reported
trial-level significance tests as new inferential evidence for P1.

\subsection{Study 2: Policy-conditioned later unaided behaviour}
\label{subsec:study-cabitza}

\citet{cabitza2026} studied 289 medical students in a simulated diagnostic
task under different AI-support policies. The published report distinguishes
AI advice alone, advice accompanied by confidence information, and a Feedback
condition providing trial-by-trial information about the correctness of both
the participant's and the AI's decisions. The authors interpret the resulting
pattern as policy-dependent diagnostic learning. The released participant-level
data allow a narrower framework-specific question to be tested directly:
whether subsequent unaided behaviour differs across interaction policies.

For P2, the focal outcome is each participant's subsequent unaided accuracy
on the 30 Phase~2 cases governed by the hidden diagnostic criterion. The
central contrasts compare Feedback with Advice and with Advice+Confidence.
The proposition does not require the behavioural difference to arise from a
particular learning mechanism; it asks whether the observed residual outcome differs across delivery policies.

An important measurement limitation is inherent to the design of the probe. Every
hidden-criterion case has ``positive'' as its correct response. Higher
accuracy on this restricted probe can therefore reflect either improved
discrimination of the hidden cue or a more liberal tendency to respond
``positive'' when only one known criterion is present. The present analysis
does not attempt to identify uniquely between those mechanisms. It treats the
outcome as an observed post-interaction behavioural probe and interprets the
between-policy contrasts as evidence of policy dependence at that level.
Assignment to conditions used a pre-specified mapping from birth month rather
than a random-number generator; consequently, causal interpretation requires
the additional assumption that this assignment was as-if random with respect
to the relevant outcomes.

\subsection{Study 3: Two-horizon generative-AI interaction policies}
\label{subsec:study-bastani}

\citet{bastani2025} assigned high-school mathematics classrooms to Control, GPT Base, or GPT Tutor. Students were randomly assigned to classrooms, whereas classrooms were allocated to arms with an integer program that matched observable characteristics under scheduling constraints; following the authors, we treat this classroom-level allocation as randomised. The public repository contains anonymised student performance data, covariates, analysis code, and timestamped student--GPT conversations \citep{bastani2025data}. The published main specification estimates intention-to-treat effects for normalised practice performance (\texttt{Part2Tot}) and a subsequent unaided exam (\texttt{Part3Tot}), with standard errors clustered at classroom level. The analysed data comprise 2{,}848 student-session observations from 44 classrooms (17 Control, 13 GPT Base, 14 GPT Tutor). The exam was taken in the same session, minutes after practice, with problems paired to the practice problems, so the second horizon provides a very short-term post-assistance performance probe.  Because allocation occurred at the classroom level, with few and unbalanced clusters and teacher fixed effects (six of the 13 GPT Base classrooms are taught by teachers with no classroom in another arm), we complement the published CR1 standard errors with CR2 standard errors and Satterthwaite degrees of freedom \citep{pustejovsky2018} and a wild cluster restricted bootstrap \citep{cameron2008,roodman2019}. The primary analyses use the authors' original sample, outcome definitions, and covariate specification; alternative specifications are reported only as sensitivity analyses. Implementation details for the small-sample inference are reported in Appendix~\ref{app:p3}.

Let $A_i\in\{0,B,T\}$ denote assignment to Control, GPT Base, or GPT Tutor. For $a\in\{B,T\}$, define the first-horizon ITT estimand
\begin{equation}
G_a^{\mathrm{ITT}}=\mathbb E\left[Y_i^{\mathrm{prac}}(a)-Y_i^{\mathrm{prac}}(0)\right],
\label{eq:bastani-G}
\end{equation}
and the second-horizon observed competence-probe estimand
\begin{equation}
\Delta O_{C,a}^{\mathrm{ITT}}=\mathbb E\left[O_{i,\mathrm{post}}^C(a)-O_{i,\mathrm{post}}^C(0)\right].
\label{eq:bastani-C}
\end{equation}

The key distinction is
\[
\Delta O_{C,a}^{\mathrm{ITT}}\neq\Delta C^{\mathrm{res}}
\]
as an identity: randomisation identifies the effect of assignment on the observed unaided competence probe, not direct observation of the latent competence state.

For P3, the analysis first reproduces the published Control-relative treatment effects and then estimates the direct Tutor--Base contrasts
\[
\Delta G_{T-B}^{\mathrm{ITT}},\qquad \Delta O_{C,T-B}^{\mathrm{ITT}}
\]
using the original clustered regression specification. This replaces descriptive subtraction of published point estimates with direct inferential contrasts.

\subsection{Study 4: Persistence across post-interaction horizons}
\label{subsec:study-corbett}

\citet{corbett2026} conducted a preregistered experiment with 375 participants holding strong psychology misconceptions. Participants received personalised Misconception AI Dialogue, Textbook Refutation, or Neutral AI Dialogue. Belief in the targeted misconception was assessed before intervention, immediately afterwards, after 10 days, and after two months. The open data and analysis materials are available on OSF \citep{corbett2026data}.

For P4, we represent the relevant observed contrast as a function of post-interaction horizon,
\begin{equation}
\Delta O_{B}^{\mathrm{res}}(\tau)=\mathbb E[O_B(\Gamma;\tau)-O_B(\Gamma^c;\tau)],
\label{eq:corbett-time}
\end{equation}
where $O_B$ is the study's belief/misconception outcome, the trajectory arguments $(\Gamma,\Gamma^c)$ are suppressed when the comparison is unambiguous, and $\tau\in\{\mathrm{post},10\mathrm{d},2\mathrm{m}\}$. The primary contrasts examine Misconception AI Dialogue relative to Textbook Refutation and Neutral AI Dialogue across all post-intervention horizons. The proposition is supported by time variation in the trajectory contrast; it does not require monotonic decay. The original study reports an immediate ANCOVA on all participants ($N=375$) and a repeated-measures ANCOVA on complete cases ($N=311$) that includes all three post-intervention horizons and a Condition $\times$ Time interaction. Our contribution is to estimate these temporal contrasts explicitly under a common specification that uses all participants, to decompose them by arm, and to examine their sensitivity to missing data. We therefore reconstructed the participant-level dataset from the released data \citep{corbett2026data}, following the authors' rules (strongest pre-intervention misconception per participant; exclusion of participants with incomplete post-test responses), reproduced both published specifications (Section~\ref{subsec:results-p4}), and re-estimated all post-interaction contrasts in a single mixed model for repeated measures (MMRM) that retains every participant with at least one post-intervention observation ($N=375$; 1{,}056 observations, including one two-month response linked to its participant through the self-reported Prolific ID because the embedded identifier was missing). The model includes condition, horizon (categorical: post, 10 days, 2 months), their interaction, and centred pre-intervention belief interacted with horizon, with an unstructured within-participant covariance matrix (generalised least squares, REML) and approximate Satterthwaite degrees of freedom (as computed by \texttt{emmeans} for \texttt{gls} fits). The central estimand for P4 is the contrast-of-contrasts
\begin{equation}
\Delta D_{M-j}(\mathrm{post},2m)=D_{M-j}(2m)-D_{M-j}(\mathrm{post}) =(M_{2m}-M_{\mathrm{post}})-(j_{2m}-j_{\mathrm{post}}), \qquad j\in\{T,N\},
\label{eq:corbett-deltaD}
\end{equation}
whose second form decomposes any change into movement of the focal and of the comparison trajectory (a descriptive decomposition of adjusted means). Holm correction is applied separately to the six horizon-specific contrasts and to the two primary temporal contrasts. Because a non-significant $\Delta D$ does not establish stability, we additionally report exploratory two one-sided tests (TOST) at several equivalence margins, which were not preregistered. Robustness is examined with a random-intercept LMM, GEE with robust standard errors, inverse-probability-of-retention weighted GEE, horizon-specific ANCOVAs on available cases, a complete-case change-score model, and a delta-adjusted multiple-imputation missing-data sensitivity analysis for departures from missing-at-random. All Study~4 analyses were run in R 4.3.3. The MMRM was fitted with nlme (version 3.1-164) using an unstructured correlation matrix and heterogeneous variances by horizon; marginal contrasts and approximate Satterthwaite degrees of freedom were obtained with emmeans (1.10.0). The random-intercept LMM used lme4 (1.1-35.1) with lmerTest (3.1-3), GEE and IPW-GEE used geepack (1.3.10), multiple imputation used mice (3.16.0), and the published repeated-measures ANCOVA was reproduced with afex (1.3-1).

\section{Results and Analysis}
\label{sec:results}

We now examine the four propositions, proposition by proposition, using the empirical mappings of Section~\ref{sec:empirical-framework}. Each subsection specifies the trajectory contrast implied by the proposition and the observable probe of the corresponding human-state component, and then distinguishes three analytical levels: the result reported in the original publication, its representation in the framework, and the framework-specific contrast estimated from the released data where the original authors did not report it. Re-expressing a published result within a common theoretical space is not equivalent to estimating a new effect.

\subsection{P1: Multidimensionality and multidirectional residual consequences}
\label{subsec:results-p1}

Proposition P1 states that a single interaction trajectory can leave
simultaneous residual consequences that differ in substantive and evaluative
meaning. \citet{vicente2025} provide an illustrative case because the
AI-supported group encountered an AI that was accurate on most training cases
but systematically incorrect on one specific class, after which both groups
completed an unaided phase.

Let $\Gamma_{\mathrm{AI}}$ denote the prior trajectory with AI advice and
$\Gamma_{\mathrm{noAI}}$ the matched unassisted trajectory. Rather than treating the
published outcomes as direct measurements of independent latent components,
we represent two observed post-interaction coordinates,
\[
\widehat{\mathbf r}_{1}
=
\left(
\widehat{\Delta O}^{\,\mathrm{res}}_{\mathrm{acc}},
\widehat{\Delta O}^{\,\mathrm{res}}_{\mathrm{bias}}
\right),
\]
where $O_{\mathrm{acc}}$ denotes an accuracy-oriented probe and
$O_{\mathrm{bias}}$ denotes later responses consistent with the AI-induced
systematic error.

During the unaided Phase~2, the previously AI-assisted group had an error
rate of approximately $.00$ on the set of cases outside the AI's systematic
error pattern, compared with approximately $.02$ in Control. Expressed as an
accuracy-oriented descriptive contrast,
\[
\widehat{\Delta O}^{\,\mathrm{res}}_{\mathrm{acc}}
\approx +.02.
\]

The magnitude is small and the outcome is close to ceiling.

The bias-related consequence is substantially larger. On the critical 40/60
cases on which the AI had systematically supplied incorrect advice during
training, the unaided Phase~2 error rate was $.26$ in the previously
AI-assisted group and $.01$ in Control,
\[
\widehat{\Delta O}^{\,\mathrm{res}}_{\mathrm{bias}}
=.26-.01=.25.
\]

The pattern also generalised to previously unseen ambiguous 50/50 cases:
participants previously exposed to the AI selected the category consistent
with the induced bias with probability $.76$, compared with $.39$ in Control,
\[
\widehat{\Delta O}^{\,\mathrm{res}}_{\mathrm{bias,gen}}
=.76-.39=.37.
\]

The case therefore illustrates why residual consequences need not admit a
single ``better'' or ``worse'' scalar description: the same prior interaction
history is associated with slightly higher observed accuracy on some case classes and
substantially worse behaviour on others. The case therefore justifies reporting several observed outcomes. Because the AI's systematic error can be represented as a shifted decision threshold, a single transmitted decision criterion could generate both consequences, so the case does not show that several latent dimensions are required, although the framework permits them.

\subsection{P2: Interaction-policy dependence}
\label{subsec:results-p2}

Proposition P2 asks whether the post-interaction human outcome depends on
\emph{how} AI capability is delivered rather than merely on whether AI is
available. \citet{cabitza2026} compared Advice, Advice+Confidence, and
trial-by-trial Feedback within the same broad diagnostic decision-support
setting.

Let
\[
P_A=\text{Advice},\qquad
P_{AC}=\text{Advice+Confidence},\qquad
P_F=\text{Feedback},
\]
and let $O_H$ denote the participant-level subsequent unaided accuracy probe
on the 30 hidden-criterion Phase~2 cases. Define
\[
\mu_p^H=\mathbb E[O_H\mid P=p].
\]

The two focal policy contrasts are
\[
D_{F-A}=\mu_F^H-\mu_A^H,
\]
and
\[
D_{F-AC}=\mu_F^H-\mu_{AC}^H.
\]

We reanalysed the released participant-level data \citep{cabitza2026data}. The three AI-supported conditions contained $N=221$ participants (Advice: $n=70$; Advice+Confidence: $n=94$; Feedback: $n=57$), all with complete responses on the 30 focal Phase~2 cases. Mean accuracy was $\widehat{\mu}_A^H=0.480$ ($SD=0.222$), $\widehat{\mu}_{AC}^H=0.462$ ($SD=0.201$) and $\widehat{\mu}_F^H=0.558$ ($SD=0.193$). Table~\ref{tab:p2-contrasts} reports the two focal contrasts. Using two-sided Welch tests, Feedback exceeded Advice by about $7.8$ percentage points (Hedges' $g=0.370$) and Advice+Confidence by about $9.6$ percentage points ($g=0.481$) on the subsequent unaided hidden-criterion probe; both contrasts remain below the $0.05$ threshold after Holm correction.

As a sensitivity analysis, an OLS model with HC3 heteroskedasticity-robust standard errors, adjusted for accuracy in the 10-case pre-intervention test phase and with unadjusted Wald inference based on a normal reference distribution, produced similar estimates (Table~\ref{tab:p2-contrasts}).

\begin{table}[ht]
\centering
\footnotesize
\setlength{\tabcolsep}{3pt}
\caption{Direct between-policy contrasts in Cabitza and Vicente on subsequent unaided accuracy for the 30 hidden-criterion Phase~2 cases (difference in proportion correct; $N=221$). Welch: two-sided Welch $t$-tests, with Holm correction across the two contrasts and Hedges' $g$. HC3-adjusted: OLS adjusted for accuracy in the 10-case pre-intervention test phase, HC3 standard errors and unadjusted Wald inference with a normal reference distribution.}
\label{tab:p2-contrasts}
\begin{tabular}{llcccccc}
\toprule
Contrast & Analysis & $\widehat D$ & $SE$ & 95\% CI & $t$ (df) & $p$ ($p_{\mathrm{Holm}}$) & $g$ \\
\midrule
Feedback $-$ Advice & Welch & 0.0779 & 0.0369 & [0.0050, 0.1508] & 2.114 (124.38) & .0365 (.0365) & 0.370 \\
 & HC3-adjusted & 0.0810 & 0.0371 & [0.0083, 0.1538] & -- & .0290 & -- \\
\addlinespace
Feedback $-$ Adv.+Conf. & Welch & 0.0958 & 0.0330 & [0.0306, 0.1611] & 2.908 (121.92) & .0043 (.0086) & 0.481 \\
 & HC3-adjusted & 0.1003 & 0.0335 & [0.0346, 0.1660] & -- & .0028 & -- \\
\bottomrule
\end{tabular}
\end{table}

These direct between-policy estimates provide evidence compatible with P2 in this context and on this probe: later unaided behaviour differs across
interaction policies. They do not uniquely identify the psychological
mechanism generating that difference. Because all hidden-criterion cases
require a positive response, higher accuracy can arise from improved
discrimination of the hidden cue, from a more liberal response criterion, or
from a combination of the two. The published table (Table~\ref{tab:p2-published}) is compatible with this ambiguity: Feedback shows higher hidden-criterion accuracy but lower accuracy
on known-criterion cases, while overall Phase~2 accuracy is similar across
policies. We therefore interpret the result as evidence of
\emph{policy-dependent residual behaviour}, not as definitive evidence that
Feedback uniquely increased latent diagnostic competence.

\begin{table}[ht]
\centering
\small
\caption{Published Phase~2 unaided accuracy in Cabitza and Vicente by case type (descriptive values reported in the original article, pooled over responses). The hidden-criterion row corresponds to the probe analysed in Section~\ref{subsec:results-p2}; the other rows provide context and are not tested here. ``Similar'' overall accuracy is a descriptive statement, not an equivalence claim. The original article reports these values truncated rather than rounded to three decimals; recomputed from the released participant-level data, the hidden-criterion means are $0.4800$, $0.4621$ and $0.5579$, which is why the Feedback mean appears as $0.558$ in Section~\ref{subsec:results-p2}.}
\label{tab:p2-published}
\begin{tabular}{lccc}
\toprule
Case type & Advice & Advice+Confidence & Feedback \\
\midrule
Hidden criterion (30 cases) & .480 & .462 & .557 \\
Known criteria (40 cases) & .783 & .801 & .742 \\
All cases (70) & .653 & .655 & .663 \\
\bottomrule
\end{tabular}
\end{table}

In framework terms,
\[ \widehat{\Delta O}_H^{\mathrm{res}} \bigl(\Gamma(P_F),\Gamma(P_A)\bigr)>0, \]
and
\[ \widehat{\Delta O}_H^{\mathrm{res}} \bigl(\Gamma(P_F),\Gamma(P_{AC})\bigr)>0, \]
where the hats denote estimated differences in observed means; reading them as trajectory effects requires the as-if-random assumption discussed in Section~\ref{subsec:study-cabitza}.

\subsection{P3: Horizon dissociation and direct AI-policy comparison}
\label{subsec:results-p3}

Proposition P3 addresses the central distinction of the framework: performance achieved while AI is available does not by itself identify the human capability observable afterwards. Bastani et al. provide a particularly informative short-horizon illustration because classroom-level assignment, which the authors treat as randomised, generates intention-to-treat contrasts both for AI-assisted practice and for a subsequent exam completed without equivalent AI support. The exam occurred within the same session, minutes after practice, and used paired problems; it should therefore be interpreted as a probe of very short-term post-assistance performance rather than durable retention.

For $a\in\{B,T\}$, where $B$ denotes GPT Base and $T$ GPT Tutor, define the empirical two-horizon coordinate
\[
\widehat{\mathbf e}^{\,\mathrm{ITT}}_a=\left(\widehat G^{\mathrm{ITT}}_a,\ \widehat{\Delta O}^{\mathrm{ITT}}_{C,a}\right).
\]

The first coordinate is the estimated effect of treatment assignment on assisted-practice performance; the second is the effect on the subsequent unaided competence probe. The latter is an observable outcome generated under the post-assistance testing condition and should not be identified with the latent competence state itself.

\paragraph{Published benchmark.} In the published main specification ($N=2{,}848$ observations), GPT Base increased normalised practice performance relative to Control by $\widehat G_B^{\mathrm{ITT}} = 0.137$ ($SE=0.031$), while its effect on the subsequent unaided exam was $\widehat{\Delta O}_{C,B}^{\mathrm{ITT}} = -0.054$ ($SE=0.022$). Hence $\widehat{\mathbf e}^{\,\mathrm{ITT}}_B = (0.137,-0.054)$
shows opposite signs of the published point estimates across the two horizons: assignment to GPT Base improved what students accomplished during assisted practice, while the estimated effect on the later unaided probe was negative. As shown below, this estimated reduction is not supported under small-sample cluster-robust inference, so it is not treated here as evidence that GPT Base harmed later performance.

GPT Tutor occupied a different position. Its effect on assisted practice was $\widehat G_T^{\mathrm{ITT}} = 0.361$ ($SE=0.032$), whereas its effect on the unaided exam was $\widehat{\Delta O}_{C,T}^{\mathrm{ITT}} = -0.004$ ($SE=0.013$), so that $\widehat{\mathbf e}^{\,\mathrm{ITT}}_T = (0.361,-0.004)$.
The second coordinate is statistically indistinguishable from Control in the published analysis; this should be described as no detectable difference on the observed exam outcome rather than as proof of equivalence.

These two vectors make P3 visible geometrically. Both AI-mediated regimes move the first coordinate in a favourable direction; the vectors describe estimated positions and, on their own, do not establish that the two policies differ at the second horizon. What they illustrate is that a favourable first coordinate does not by itself identify the second.
The inferential weight of this distinction is assessed below with small-sample corrections; performance-only evaluation would not raise the question at all.
\paragraph{Reproduction check.} We first reproduced the original main ITT specification from the released microdata and analysis script \citep{bastani2025data}. Following the authors' code, honours-designated classrooms were excluded and the two outcomes $Y\in\{\texttt{Part2Tot},\texttt{Part3Tot}\}$ were modelled separately as
\[
Y \sim \texttt{GPTBase}+\texttt{GPTTutor}+\texttt{gpa\_prev} +\texttt{teacher}+\texttt{Session}+\texttt{Grader}+\texttt{Year},
\]
with cluster-robust covariance at the classroom level ($N=2{,}848$ observations, 44 classroom clusters). The reproduced treatment coefficients agree with the published benchmarks: for assisted practice, $\widehat\beta_B = 0.1370$ ($SE=0.0310$) and $\widehat\beta_T = 0.3613$ ($SE=0.0317$); for the subsequent unaided exam, $\widehat\beta_B = -0.0543$ ($SE=0.0224$) and $\widehat\beta_T = -0.0043$ ($SE=0.0134$). This reproduction recovers the published values to rounding and therefore provides a check that the framework-specific contrasts are being computed from the intended specification.

\paragraph{Direct Tutor--Base contrasts.} The CR1 results in this paragraph are reported as the conventional reference; the interpretation of P3 rests on the CR2/Satterthwaite and wild-bootstrap results that follow. The published Control-relative coefficients imply point differences of $0.3613-0.1370=0.2243$ for assisted practice and $-0.0043-(-0.0543)=0.0500$ for the subsequent unaided exam. Their uncertainty cannot be obtained by treating the separately reported standard errors as independent, so we tested $H_0:\beta_T-\beta_B=0$ directly from the cluster-robust covariance matrix of each original regression, with $df=43$. On assisted practice, $\Delta\widehat G_{T-B}^{\mathrm{ITT}} = 0.2243$ ($SE=0.0342$, 95\% CI $[0.1553,0.2932]$, $t(43)=6.561$, $p<.001$): assignment to GPT Tutor produced a substantially larger observed assisted-practice gain than assignment to GPT Base. On the subsequent unaided exam, $\Delta\widehat O_{C,T-B}^{\mathrm{ITT}} = 0.0500$ ($SE=0.0233$, 95\% CI $[0.0030,0.0970]$, $t(43)=2.146$, $p=.0375$).
\paragraph{Small-sample inference.} Conventional inference combines CR1 standard errors with a $t_{43}$ reference distribution. The design warrants caution about this reference: treatment varies across only 44 unbalanced classroom clusters, and with teacher fixed effects six of the 13 GPT Base classrooms are taught by teachers with no classroom in another arm, so they provide no within-teacher comparison between treatments, although they still contribute to the estimation of other model parameters. CR2 standard errors combined with Satterthwaite degrees of freedom \citep{pustejovsky2018} adjust both the covariance estimate and the reference distribution to the design and to each contrast, giving effective degrees of freedom of about 10--14 here, and a wild cluster restricted bootstrap \citep{cameron2008} gives similar results (Table~\ref{tab:p3-inference}). The six $p$-values are reported without multiplicity adjustment because they correspond to distinct estimands at two horizons; conclusions rest primarily on interval estimates and are unchanged under a Holm correction across all six contrasts (practice: CR2 $p_{\mathrm{Holm}}\le .028$, bootstrap $p_{\mathrm{Holm}}\le .018$; exam: CR2 $p_{\mathrm{Holm}}\ge .33$, bootstrap $p_{\mathrm{Holm}}\ge .31$).

\begin{table}[t]
\centering
\small
\caption{Inference for the Study~3 contrasts in Bastani et al.\ under the published CR1 standard errors and under small-sample corrections. Estimates are identical across columns; only the uncertainty differs. $p_{\mathrm{CR1}}$ uses a $t_{43}$ reference distribution, which is more conservative than the normal reference implied by the original table (GPT Base on the exam: normal, $p=.015$; $t_{43}$, $p=.020$). $p_{\mathrm{WCR}}$: wild cluster restricted bootstrap, $B=9{,}999$. The CR2 Satterthwaite degrees of freedom depend only on the design matrix, the clustering and the contrast, not on the outcome, and are therefore identical at the two horizons.}
\label{tab:p3-inference}
\begin{tabular}{llrrrrr}
\toprule
Horizon & Contrast & Estimate & $p_{\mathrm{CR1}}$ & 95\% CI (CR2) & df & $p_{\mathrm{CR2}}$ / $p_{\mathrm{WCR}}$ \\
\midrule
Practice & Base $-$ Control  & $0.137$  & $<.001$ & $[0.046,\,0.228]$  & 10.9 & $.007$ / $.003$ \\
Practice & Tutor $-$ Control & $0.361$  & $<.001$ & $[0.275,\,0.447]$  & 13.8 & $<.001$ / $<.001$ \\
Practice & Tutor $-$ Base    & $0.224$  & $<.001$ & $[0.124,\,0.324]$  & 9.9  & $<.001$ / $.005$ \\
\addlinespace
Exam & Base $-$ Control  & $-0.054$ & $.020$ & $[-0.124,\,0.015]$ & 10.9 & $.112$ / $.104$ \\
Exam & Tutor $-$ Control & $-0.004$ & $.749$ & $[-0.042,\,0.033]$ & 13.8 & $.808$ / $.781$ \\
Exam & Tutor $-$ Base    & $0.050$  & $.038$ & $[-0.020,\,0.120]$ & 9.9  & $.144$ / $.131$ \\
\bottomrule
\end{tabular}
\end{table}

Three conclusions follow. First, every assisted-practice effect is robust. Second, the two unaided-exam contrasts that are nominally significant under CR1 (GPT Base versus Control and Tutor--Base) are not detected under either CR2/Satterthwaite inference or the wild cluster restricted bootstrap. Both estimates are also sensitive to the fixed-effect specification: without teacher fixed effects the Tutor--Base exam contrast falls to $0.026$ ($p=.20$) and the GPT Base exam effect to $-0.020$ ($p=.35$) (Table~\ref{tab:app-p3-fe}). Third, the GPT Tutor exam effect is estimated close to zero: its CR2 95\% CI $[-0.042,0.033]$ is compatible with a small negative or small positive effect and excludes gains larger than $0.033$ on the normalised grade scale (0--1, share of the maximum rubric score). This is a bound on plausible effects, not evidence of an exactly null effect.

This refines the evidential basis for P3. The robust result is neither that GPT Base harmed later performance nor that GPT Tutor outperformed GPT Base after assistance was removed, but that GPT Tutor produced the largest assisted-practice gain in the study ($0.361$ relative to Control) while no corresponding difference was detected on the subsequent unaided exam ($-0.004$, CR2 95\% CI $[-0.042,0.033]$). This is an assisted improvement with no detected later difference, not evidence of unchanged competence or statistical equivalence; classifying it as the preservation regime of Section~\ref{subsec:signatures} would require a justified equivalence margin for the exam scale, which is not available. The direct Tutor--Base comparison points the same way: the two policies differ robustly during assisted practice, whereas the available number of classrooms does not support a corresponding difference on the subsequent unaided probe. The result therefore provides evidence for the empirical non-identification claim associated with P3: a clear assisted gain is insufficient to establish a positive residual competence effect. It does so without testing the structural possibility in Equation~\eqref{eq:horizon-proposition} and without establishing that learning failed to occur.

\subsection{P4: Persistence, decay, and comparator-dependent temporal evolution}
\label{subsec:results-p4}

Proposition P4 adds a temporal dimension to the residual contrast. A difference detected immediately after interaction need not have the same magnitude at a later horizon, and the direction of that evolution must be defined relative to a specified comparison trajectory. \citet{corbett2026} provide a particularly informative illustration because belief in the targeted misconception was assessed before intervention, immediately afterwards, after 10 days, and after two months.

Let $M$ denote personalised Misconception AI Dialogue, $T$ Textbook Refutation, and $N$ Neutral AI Dialogue, and define $D_{M-j}(\tau)=\mathbb E[O_B\mid M,\tau]-\mathbb E[O_B\mid j,\tau]$, $j\in\{T,N\}$, so that negative values favour Misconception AI Dialogue. Note that the Textbook Refutation passages were generated to address the misconception indirectly, without explicitly naming or refuting it \citep{corbett2026}; $D_{M-T}$ therefore contrasts personalised dialogue with accurate expository text rather than with a classical refutation text.

\paragraph{Reproduction of the published benchmark.} From the released data we reproduced the published immediate contrasts exactly ($D_{M-N}(\mathrm{post})=-36.55$, $SE=3.43$; $D_{M-T}(\mathrm{post})=-9.77$, $SE=3.43$; Tukey-adjusted $p<.001$ and $p=.013$) and the published complete-case repeated-measures ANCOVA ($N=311$): Condition $\times$ Time $F(4,614)=6.63$, $p<.001$, $\eta_p^2=.041$, with $D_{M-N}=-29.09$ and $D_{M-T}=-12.17$ at 10 days and $D_{M-N}=-19.95$ and $D_{M-T}=-8.11$ (Tukey $p=.098$) at two months. The published main effect of Time ($F=0.28$, $p=.756$) is also reproduced exactly when the authors' specification is used (Type~III tests with treatment contrasts and an uncentred covariate). Under that coding, however, the term is the simple effect of Time in the Misconception AI arm at a pre-intervention belief of zero rather than an average over conditions; with a centred covariate and sum-to-zero contrasts it is $F(2,614)=2.54$, $p=.080$. This does not affect the contrasts used here. In our reconstruction, the value reported in the original article as the immediate effect of intervention, $F(2,371)=46.59$, corresponds to the omnibus model $F(3,371)$; the covariate-adjusted effect of condition is $F(2,371)=60.56$, $p<.001$, $\eta_p^2=.246$.\footnote{Our reconstruction indicates that the reported $\eta_p^2=.230$ corresponds to condition entered before the covariate (sequential sums of squares) and is therefore not adjusted for pre-intervention belief.} Two features of the published benchmarks matter for P4. First, the contrasts highlighted in the article combine the full sample (immediate) and the complete-case sample (follow-ups). Within the complete-case model itself the immediate contrasts are $-34.68$ ($M$--$N$) and $-8.67$ ($M$--$T$), so the implied changes to two months are about $14.7$ and $0.6$ points, but these changes are not tested there as comparator-specific contrasts. Second, the statement that the $M$--$T$ difference was ``no longer statistically reliable'' at two months rests on a Tukey-adjusted test in the complete-case sample; the unadjusted value is $p=.039$, and the point estimate ($-8.11$) is close to the immediate one.

\paragraph{Attrition.} Retention did not differ reliably by condition at 10 days (Neutral AI $93.6\%$, Textbook $98.4\%$, Misconception AI $92.8\%$; $\chi^2(2)=4.77$, $p=.092$) but did at two months ($80.0\%$, $92.0\%$, $88.0\%$; $\chi^2(2)=8.08$, $p=.018$; Table~\ref{tab:p4-attrition}).\footnote{The reconstructed follow-up samples ($n=356$ at 10 days and $n=325$ at two months) differ from the 359 and 326 reported in the original article, which count all rows in the follow-up files. At 10 days the difference corresponds to one Misconception AI response whose Prolific ID does not appear in the first-session file and two unfinished responses without the outcome item; at two months it corresponds to one Misconception AI response whose Prolific ID does not appear in the first-session file. One completed two-month response (Neutral AI) lacks the embedded Prolific ID; it is linked to its participant through the self-reported ID and included in our analyses, but set aside when reproducing the published complete-case benchmark ($N=311$). Two retained participants (one Neutral AI, one Misconception AI) have a highest pre-intervention rating of exactly 50, below the published inclusion threshold (${>}50$); they are kept so that the analysed sample matches the authors' (125 per arm).} In a logistic model (Table~\ref{tab:app-p4-retention}), retention was associated with condition, whereas no association was detected with pre-intervention ($p=.19$) or post-intervention belief ($p=.17$). This does not verify that missingness is ignorable: the MMRM and the weighted analyses rely on missingness being at random given the observed variables, an assumption that cannot be tested from the observed data. The delta-adjusted missing-data sensitivity analysis below examines departures from it, and differential two-month attrition remains a limitation.

\paragraph{Common longitudinal specification.} In the MMRM using all 375 participants, the Condition $\times$ Time interaction was $F(4,348.7)=6.90$, $p<.001$. Table~\ref{tab:p4-contrasts} reports the six horizon-specific residual contrasts and the direct temporal contrasts.

\begin{table}[t]
\centering
\small
\caption{Residual belief contrasts for Misconception AI Dialogue estimated in a single MMRM ($N=375$; unstructured covariance, baseline $\times$ horizon, approximate Satterthwaite df). Negative $D$ favours Misconception AI. $\Delta D$ is the change in the contrast between the immediate and two-month horizons (Eq.~\ref{eq:corbett-deltaD}). Holm correction is applied within the six horizon-specific contrasts and within the two temporal contrasts.}
\label{tab:p4-contrasts}
\begin{tabular}{llrcrr}
\toprule
Comparator & Estimand & Estimate & 95\% CI & $p$ & $p_{\mathrm{Holm}}$ \\
\midrule
Neutral AI ($N$) & $D_{M-N}(\mathrm{post})$ & $-36.55$ & $[-43.30,\,-29.80]$ & $<.001$ & $<.001$ \\
 & $D_{M-N}(10d)$ & $-29.63$ & $[-36.90,\,-22.37]$ & $<.001$ & $<.001$ \\
 & $D_{M-N}(2m)$ & $-20.77$ & $[-28.24,\,-13.31]$ & $<.001$ & $<.001$ \\
 & $\Delta D_{M-N}(\mathrm{post},2m)$ & $15.78$ & $[8.93,\,22.63]$ & $<.001$ & $<.001$ \\
\addlinespace
Textbook ($T$) & $D_{M-T}(\mathrm{post})$ & $-9.77$ & $[-16.51,\,-3.03]$ & $.005$ & $.009$ \\
 & $D_{M-T}(10d)$ & $-12.90$ & $[-20.11,\,-5.69]$ & $<.001$ & $.001$ \\
 & $D_{M-T}(2m)$ & $-8.21$ & $[-15.53,\,-0.90]$ & $.028$ & $.028$ \\
 & $\Delta D_{M-T}(\mathrm{post},2m)$ & $1.56$ & $[-5.13,\,8.24]$ & $.647$ & $.647$ \\
\bottomrule
\end{tabular}
\end{table}

\paragraph{Direct temporal contrast.} Relative to Neutral AI Dialogue, the residual contrast shrank by $\Delta\widehat D_{M-N}(\mathrm{post},2m)=15.78$ points (95\% CI $[8.93,22.63]$, $p_{\mathrm{Holm}}<.001$). The decomposition in Eq.~\ref{eq:corbett-deltaD} shows that this change had two sources: belief in the Misconception AI arm rebounded by $6.01$ points (95\% CI $[1.25,10.77]$), while belief in the Neutral AI arm \emph{declined} by $9.77$ points (95\% CI $[4.84,14.69]$) in the absence of any corrective content. In this descriptive decomposition of adjusted observed means, approximately 62\% ($9.77/15.78$) of the attenuation of $D_{M-N}$ corresponds to movement of the comparison trajectory; it does not identify individual-level mechanisms of change. This share also assumes that two-month data are missing at random. In the delta-adjusted analysis described below, if Neutral participants lost to follow-up held two-month beliefs 20 or 30 points higher than imputed, the Neutral share falls to 49\% and 39\% under the primary imputation model, which is unbounded on the 0--100 scale, and to 56--57\% and 53--54\% when imputed values are restricted to that scale (Appendix~\ref{app:p4}). Only upward departures, which reduce the Neutral share, were examined; under scale-respecting imputation the comparison arm accounts for more than half of the attenuation in every scenario. Part of the Neutral decline may also reflect regression to the mean, because the targeted item was selected as each participant's highest pre-intervention rating. Relative to Textbook Refutation, the change in the contrast was small: $\Delta\widehat D_{M-T}(\mathrm{post},2m)=1.56$ (95\% CI $[-5.13,8.24]$, $p=.647$), because both corrective arms rebounded by similar amounts ($M$: $6.01$; $T$: $4.45$, 95\% CI $[-0.25,9.15]$). The 95\% CI alone does not establish stability. In an exploratory equivalence analysis (margins chosen after the data were available and not preregistered), the 90\% CI for $\Delta D_{M-T}$ ($[-4.05,7.16]$) lies within $\pm7.5$ points (TOST $p=.041$) and $\pm10$ points ($p=.007$), but not within $\pm5$ points ($p=.156$). Because a margin of $\pm7.5$ points would treat as equivalent a change of about three quarters of the immediate Misconception AI advantage over Textbook Refutation ($9.8$ points), no single margin is privileged: the three results are reported together as an exploratory sensitivity analysis, and the primary conclusion is that no change in the $M$--$T$ contrast was detected. At two months, Misconception AI remained lower than Textbook Refutation by $8.21$ points ($p_{\mathrm{Holm}}=.028$), although the detectability of this horizon-specific contrast is sensitive to specification (see below).

\paragraph{Robustness.} The temporal contrast estimates were similar across specifications (Table~\ref{tab:p4-robustness}): $\Delta D_{M-N}$ ranged from $14.80$ (complete-case change-score model) to $15.80$ (LMM), and $\Delta D_{M-T}$ from $0.56$ to $2.30$, with every 95\% CI for $\Delta D_{M-N}$ excluding zero and every CI for $\Delta D_{M-T}$ including it. The pattern of detected and undetected changes held for GEE with exchangeable working correlation and for inverse-probability-weighted GEE, with retention weights estimated from condition, pre- and post-intervention belief. The horizon-specific contrast $D_{M-T}(2m)$ was consistent in size ($-7.47$ to $-8.21$) but its significance depended on the specification ($p$ from $.025$ to $.080$). In the delta-adjusted missing-data sensitivity analysis (Table~\ref{tab:app-p4-tip}), shifting the imputed two-month values of any single arm by up to $\pm30$ points, a range chosen because it corresponds to roughly one pooled standard deviation of two-month belief (about 29.8 points) and is about three times the largest estimated within-arm change between post-test and two months (9.8 points) preserved the detected attenuation relative to Neutral ($\Delta D_{M-N}$ estimates $9.56$--$21.64$, all 95\% CIs excluding zero) and the absence of a detected change relative to Textbook ($\Delta D_{M-T}$ estimates $-2.16$ to $5.08$, all 95\% CIs including zero). Equivalence was not re-evaluated under these scenarios, so this robustness applies to the absence of a detected change, not to equivalence. The same pattern held when imputations were restricted to the 0--100 belief scale (Appendix~\ref{app:p4}).

\paragraph{Interpretation.} There is no single answer to whether the AI-induced advantage ``persisted.'' Relative to Neutral AI Dialogue, the residual contrast changed materially over two months, and most of that change corresponds, in adjusted means, to movement of the comparison trajectory: about 62\% if missingness is at random and 53--64\% across Neutral-arm shifts from $0$ to $+30$ points when imputations respect the 0--100 scale (39--62\% with unbounded imputation). The attenuation should therefore not be read as a 15.8-point loss of the improvement within the intervened group. Relative to Textbook Refutation, the horizon-specific contrasts remained of broadly comparable magnitude and no change was detected, although equivalence was not established at $\pm5$ points and the two-month significance depends on the specification. The published impression of ``convergence'' between the two corrective conditions is thus compatible with reduced precision in the smaller complete-case sample and multiplicity adjustment rather than with a demonstrated narrowing of the contrast. Persistence is both \emph{time-relative} and \emph{comparison-relative}: $\Delta O_B^{\mathrm{res}}(\Gamma,\Gamma^c;\tau)$ cannot be interpreted without specifying both $\Gamma^c$ and $\tau$, and a change in a residual contrast can originate in the comparison trajectory as much as in the focal one. These results concern the adjusted observed belief contrast, not a direct measurement of participants' latent belief state.

\begin{table}[ht]
\centering
\small
\caption{Observed outcomes by condition and horizon in the reconstructed Corbett--Tangen sample. Retention differed by condition at two months ($\chi^2(2)=8.08$, $p=.018$) but not at 10 days ($\chi^2(2)=4.77$, $p=.092$).}
\label{tab:p4-attrition}
\begin{tabular}{lrrrrrrr}
\toprule
Condition & $n_{\mathrm{pre}}$ & $n_{\mathrm{post}}$ & $n_{10d}$ & $n_{2m}$ & Complete & Ret.\ 10d & Ret.\ 2m \\
\midrule
Neutral AI & 125 & 125 & 117 & 100 & 95 & 93.6\% & 80.0\% \\
Textbook & 125 & 125 & 123 & 115 & 113 & 98.4\% & 92.0\% \\
Misconception AI & 125 & 125 & 116 & 110 & 104 & 92.8\% & 88.0\% \\
\bottomrule
\end{tabular}
\end{table}

\begin{table*}[ht]
\centering
\small
\caption{Sensitivity of the Study~4 residual contrasts to the longitudinal specification. Entries are point estimates; 95\% CIs are given in brackets for the temporal contrasts $\Delta D$, and $p$-values for $D_{M-T}(2m)$, whose significance varies across specifications. Confidence intervals for all horizon-specific entries are reported in Table~\ref{tab:app-p4-ci}. MMRM: primary model ($N=375$). LMM: random intercept, condition $\times$ horizon with centred baseline as a main effect only, Kenward--Roger df. GEE: exchangeable working correlation, robust SEs. IPW-GEE: independence working correlation with inverse-probability-of-retention weights. ANCOVA: separate horizon-specific models on available cases; the $\Delta D$ entry in this column is a complete-case change-score model ($N=312$).}
\label{tab:p4-robustness}
\begin{tabular}{lccccc}
\toprule
Estimand & MMRM & LMM & GEE & IPW-GEE & ANCOVA \\
\midrule
$D_{M-N}(\mathrm{post})$ & $-36.55$ & $-36.61$ & $-36.55$ & $-36.55$ & $-36.55$ \\
$D_{M-N}(10d)$ & $-29.63$ & $-29.52$ & $-29.59$ & $-29.51$ & $-29.02$ \\
$D_{M-N}(2m)$ & $-20.77$ & $-20.81$ & $-20.83$ & $-20.78$ & $-20.22$ \\
$D_{M-T}(\mathrm{post})$ & $-9.77$ & $-9.73$ & $-9.77$ & $-9.77$ & $-9.77$ \\
$D_{M-T}(10d)$ & $-12.90$ & $-12.96$ & $-12.93$ & $-13.15$ & $-12.76$ \\
$D_{M-T}(2m)$ & $-8.21$ & $-8.20$ & $-8.20$ & $-7.47$ & $-7.53$ \\
 & {\scriptsize $p=.028$} & {\scriptsize $p=.025$} & {\scriptsize $p=.044$} & {\scriptsize $p=.080$} & {\scriptsize $p=.050$} \\
\addlinespace
$\Delta D_{M-N}(\mathrm{post},2m)$ & $15.78$ & $15.80$ & $15.72$ & $15.77$ & $14.80$ \\
 & {\scriptsize $[8.93,22.63]$} & {\scriptsize $[9.54,22.05]$} & {\scriptsize $[9.02,22.41]$} & {\scriptsize $[8.80,22.73]$} & {\scriptsize $[7.54,22.07]$} \\
$\Delta D_{M-T}(\mathrm{post},2m)$ & $1.56$ & $1.53$ & $1.57$ & $2.30$ & $0.56$ \\
 & {\scriptsize $[-5.13,8.24]$} & {\scriptsize $[-4.57,7.62]$} & {\scriptsize $[-5.64,8.78]$} & {\scriptsize $[-5.18,9.78]$} & {\scriptsize $[-6.39,7.51]$} \\
\bottomrule
\end{tabular}
\end{table*}

\subsection{Cross-study synthesis}
\label{subsec:cross-study}

The four studies should not be read as replications of one psychological effect. They differ in domain, intervention, state component, measurement scale, and temporal design. Their value lies instead in examining four structurally distinct implications of the same trajectory-based representation, with P1 serving as an illustration and P2--P4 as inferential contrasts. Table~\ref{tab:cross-synthesis} summarises the empirical correspondence.

\begin{table*}[!htbp]
\centering
\footnotesize
\renewcommand{\arraystretch}{1.15}
\caption{Cross-study synthesis of the four framework propositions. Numerical entries distinguish published benchmarks from framework-specific secondary contrasts estimated from the released microdata.}
\label{tab:cross-synthesis}
\begin{tabularx}{\textwidth}{
>{\raggedright\arraybackslash}p{0.13\textwidth} >{\raggedright\arraybackslash}p{0.08\textwidth} >{\raggedright\arraybackslash}p{0.18\textwidth} >{\raggedright\arraybackslash}p{0.23\textwidth} >{\raggedright\arraybackslash}X}
\toprule
\textbf{Property} & \textbf{Study} & \textbf{Published empirical pattern} & \textbf{Framework representation} & \textbf{What the case adds} \\
\midrule

P1\newline Multi\-dimen\-sionality
&
Vicente et al.
&
Published unaided Phase~2 outcomes show a small descriptive accuracy difference on some
case classes but large reproduction and generalisation of the AI-induced
systematic bias on others.
&
Two observed residual coordinates,
$\widehat{\mathbf r}_1=
(\widehat{\Delta O}_{acc}^{\mathrm{res}},
\widehat{\Delta O}_{bias}^{\mathrm{res}})$,
without assuming independent latent components.
&
One interaction history can have residual consequences with different
evaluative directions; a common transmitted decision criterion remains a
plausible explanation. \\
\addlinespace
P2\newline Policy dependence
&
Cabitza \& Vicente
&
Feedback is associated with higher subsequent unaided accuracy on the
hidden-criterion probe than Advice or Advice+Confidence.
&
Direct secondary contrasts:
$\widehat D_{F-A}=0.0779$
(95\% CI $[0.0050,0.1508]$);
$\widehat D_{F-AC}=0.0958$
(95\% CI $[0.0306,0.1611]$).
&
Later unaided behaviour differs across interaction policies; the restricted
probe does not uniquely distinguish improved discrimination from a shifted
response criterion.
\\

\addlinespace

P3\newline Horizon dissociation & Bastani et al. & GPT Base $(0.137,-0.054)$ and GPT Tutor $(0.361,-0.004)$ occupy different positions in assisted-practice $\times$ unaided-exam space. & $\widehat{\mathbf e}^{\,\mathrm{ITT}}_a =(\widehat G_a^{\mathrm{ITT}}, \widehat{\Delta O}_{C,a}^{\mathrm{ITT}})$; {Tutor--Base contrasts with CR2 small-sample inference: \newline $\Delta\widehat G^{\mathrm{ITT}}_{T-B}=0.224$ (95\% CI $[0.124,0.324]$), \newline $\Delta\widehat O^{\mathrm{ITT}}_{C,T-B}=0.050$ (95\% CI $[-0.020,0.120]$).}& {Current augmentation does not identify the later observed capability probe: the largest assisted gain (GPT Tutor, $+0.36$ on the 0--1 scale) coexists with no detected unaided difference. The two policies differ detectably only at the assisted horizon.} \\

\addlinespace

P4\newline Temporal persistence & Corbett \& Tangen & {Personalised AI dialogue initially outperformed both comparators; the published complete-case analysis reported a difference relative to Neutral AI at two months and a non-significant (Tukey-adjusted) difference relative to Textbook Refutation.} & {Common MMRM ($N=375$); direct temporal contrasts: \newline $\Delta\widehat D_{M-N}(\mathrm{post},2m)=15.78$ (95\% CI $[8.93,22.63]$), \newline $\Delta\widehat D_{M-T}(\mathrm{post},2m)=1.56$ (95\% CI $[-5.13,8.24]$).} & {A residual contrast is not temporally fixed, and its change can come from the comparator: the $M$--$N$ contrast shrank largely through movement of the Neutral trajectory under the missing-at-random assumption (descriptive decomposition of adjusted means), whereas no change from post-test to two months was detected in the $M$--$T$ contrast.} \\

\bottomrule
\end{tabularx}
\end{table*}

Several conclusions follow from the joint pattern. First, residual
consequences need not be scalar. The Vicente case illustrates how one prior
interaction history can be associated with slightly higher observed accuracy on some case classes while transmitting a decision rule that substantially
worsens performance on others. This does not require the two consequences to
represent independent latent-state changes. Second, the relevant exposure
cannot be reduced to binary AI access. In Cabitza and Vicente, subsequent
unaided behaviour differs directly across support policies: Feedback exceeds
both Advice and Advice+Confidence on the hidden-criterion accuracy probe,
although the restricted outcome does not uniquely identify whether the
difference reflects improved discrimination or a response-criterion shift.
Third, contemporaneous success should not be assumed to be a proxy for later human
capability. In Bastani et al., no difference from Control was detected on the
paired unaided exam for the policy producing the largest assisted gain, and
the nominal GPT Base reduction and Tutor--Base exam advantage do not survive
small-sample cluster-robust inference. Fourth, residual contrasts are
themselves dynamic and comparison-relative. In the Corbett--Tangen data, the contrast with Neutral AI shrank substantially between the immediate and two-month horizons, with about 62\% of that attenuation corresponding to movement of the Neutral comparison trajectory under missingness at random and 53--64\% across the scale-respecting sensitivity scenarios (39--62\% with unbounded imputation), whereas no change between the immediate and two-month assessments was detected in the contrast with Textbook Refutation (Section~\ref{subsec:results-p4}).

The resulting empirical picture is therefore more specific than the general claim that ``AI changes humans.'' Across the four empirical settings, the framework isolates four distinct properties of that change:
\[
\begin{aligned}
\text{P1:}\;&\quad \text{which dimensions change?}\\
\text{P2:}\;&\quad \text{which interaction policies produce the change?}\\
\text{P3:}\;&\quad \text{does assisted success predict what remains afterwards?}\\
\text{P4:}\;&\quad \text{how does the trajectory contrast evolve over time?}
\end{aligned}
\]

Together, these questions convert the residual human state from a purely conceptual construct into a framework that can be partially operationalised through observable, domain-specific probes. The studies do not estimate a common latent variable or validate every component of $\X$; they show that the framework's central distinctions can be populated by observable quantities in existing experimental designs.

\FloatBarrier
\section{Discussion}
\label{sec:discussion}

The analyses support a broader change in the evaluation boundary of Human--AI Interaction. The relevant question is not only what a human--AI configuration achieves while assistance is available, but also whether different interaction histories leave different human states afterwards.

The implications are therefore primarily structural. The Residual Human States framework does not prescribe independence from AI, nor does it treat every post-interaction difference as learning, harm, or deskilling. It provides a framework for asking when later human capability matters, how interaction policy may shape that capability, which comparison trajectory is relevant, and how any resulting difference evolves over time.

\subsection{From AI performance to trajectory-based evaluation}

The Residual Human States framework adds a second evaluative object to HAI without replacing established measures of model capability, joint performance, trust, reliance, usability, or immediate task success:
\[
\text{what AI enables now}
\qquad+\qquad
\text{what humans carry forward}.
\]

The second term matters when humans are expected to interpret, challenge, transfer, recover, or continue acting without equivalent assistance. In such settings, evaluating only contemporaneous performance leaves part of the coupled system's longer-term performance envelope unobserved.

This distinction also separates residual-state analysis from simple AI-withdrawal comparisons. Withdrawal can reveal what a person can do once assistance is removed, but a performance drop does not establish that prior AI use degraded the human. Functional dependence on a useful tool and trajectory-induced human change are different phenomena. A residual effect requires comparison with a relevant alternative trajectory:
\[
\Delta \X^{\mathrm{res}}(\Gamma,\Gamma^c;\tau).
\]

Accordingly, the framework does not privilege unaided performance as intrinsically superior. It asks whether alternative interaction histories produce meaningfully different later human states.

\subsection{Interaction policy as a trajectory-shaping variable}

The interaction policy $P_t$ is not merely an interface choice. It determines when the human reasons, when AI intervenes, which errors or disagreements become visible, and whether feedback is processed as information for future action or consumed only as a current solution. Comparable AI capability can therefore generate different human trajectories when delivered through answer-first assistance, confidence cues, critique, progressive hints, or corrective feedback.

The empirical evidence supports this distinction at the structural level. The diagnostic case shows that later unaided behaviour differs across assistance policies, although the available outcome cannot determine whether the difference reflects improved discrimination, a shifted response criterion, or both. The mathematics case likewise shows that policies can be clearly separated by assisted performance without an equally clear separation on the subsequent unaided probe. The robust conclusion is therefore not that one policy universally produces more learning, but that different delivery policies can be associated with different later outcomes, which makes them candidates for prospective manipulation as causal design variables.

\subsection{Human agency and governable dependence}

Residual human-state analysis is compatible with useful offloading and deliberate delegation. Offloading can improve performance \citep{risko2016,clark2025}, reliance should be calibrated rather than eliminated \citep{lee2004,eckhardt2025}, and repeated dependence on AI need not be intrinsically maladaptive \citep{herrera2026,mladin2026}. The objective is therefore not maximal human independence, but an appropriate distribution of capability across the human--AI system.

Which capabilities should remain human is role-dependent. Some functions may be delegated with little consequence; others must remain sufficiently available because humans are expected to interpret, contest, supervise, recover, or assume responsibility. A person may therefore remain formally ``in the loop'' while the practical capacity to exercise those functions changes over time.

This motivates the idea of \emph{governable dependence}: dependence that is observable, role-appropriate, and compatible with preservation or development of those human capacities that remain functionally necessary. The Residual Human States framework does not determine which capacities should be preserved; it makes their evolution an explicit empirical question.

\subsection{Limitations and future research}

The present evidence has clear inferential boundaries. None of the four studies was originally designed to validate the Residual Human States framework, so the empirical analyses provide structured secondary evidence rather than prospective validation of a complete latent-state model. The studies also differ in domain, outcome, intervention, and horizon; their results are structurally comparable but not metrically commensurable. Observable outcomes should therefore be interpreted as probes of selected human-state dimensions rather than direct measurements of the full state vector $\X_t$. Because the four cases were purposively selected to illustrate the propositions, they do not indicate how prevalent these properties are in Human--AI Interaction research.

The empirical depth is intentionally asymmetric. Study~1 is primarily illustrative: its multidirectional outcomes may arise from a common transmitted decision criterion and do not establish distinct latent-state changes. Study~2 provides stronger evidence of policy-dependent later behaviour, but its restricted accuracy probe cannot separate discrimination from response criterion, and causal interpretation depends on the plausibility of the birth-month assignment as effectively random. Because the two focal comparisons and the hidden-criterion probe were chosen knowing that the original article reported differences there, this analysis should be interpreted as a targeted secondary analysis rather than as a fully confirmatory test. Study~3 benefits from classroom-level allocation that the original authors treat as randomised but is constrained by only 44 unequally allocated clusters and limited within-teacher treatment variation; its interpretation therefore rests on small-sample robust inference rather than conventional CR1 significance. Moreover, its unaided outcome is measured minutes after assisted practice and should not be interpreted as evidence of durable retention. Study~4 provides the clearest evidence about temporal evolution, but longitudinal inference remains dependent on unverifiable assumptions about missingness despite the similarity of the main temporal contrasts across sensitivity analyses.

Several substantive questions consequently remain open. First, the present cases identify residual differences more clearly than the mechanisms producing them. Future work should manipulate interaction trajectories prospectively while holding AI capability as constant as possible, allowing the causal effects of timing, feedback, disagreement, explanation, or reconstruction to be isolated. Second, later human outcomes should be observed across multiple horizons so that immediate transfer, delayed retention, attenuation, and recovery are not conflated. Third, domain-specific measurement should separate latent mechanisms where aggregate performance is ambiguous, including the distinction between discrimination and decision criterion in diagnostic settings. Fourth, comparing the same type of interaction policy across tasks and domains under a common design would make it possible to test whether residual effects depend on domain, a question that the present cross-study comparison cannot answer.

A further priority is heterogeneity. The baseline human state $\X_0$ is likely to moderate the effects of assistance: expertise, prior knowledge, reliance propensity, and metacognitive calibration may change which interaction policy is beneficial and what form residual change takes. Repeated interaction also raises a deeper trajectory question: residual effects may accumulate, reverse, or become path-dependent as earlier AI exposure changes how later assistance is processed.

The next empirical stage should therefore move from documenting residual states to identifying the interaction mechanisms that generate them. The central design question becomes not simply whether AI should assist, but \emph{which assistance policy produces the desired balance between current human--AI performance and the human capabilities required afterwards}. That question connects residual-state evaluation directly to the future design of adaptive Human--AI systems.

\section{Conclusions}
\label{sec:conclusion}

Human--AI Interaction should not be evaluated only while assistance is present. AI can improve current decisions, accelerate task completion, or increase assisted performance while also changing what the human will later know, notice, trust, contest, or be able to do. When future human judgment remains functionally relevant, these post-interaction consequences are part of the system being evaluated.

The Residual Human States framework provides a common representation for this problem. It treats the human state produced by an interaction history as a distinct evaluative object, defined relative to a relevant alternative trajectory, assessed through domain-appropriate observable probes, and indexed by post-interaction horizon. The framework does not assume that AI assistance is intrinsically beneficial or harmful, nor that unaided performance should always be privileged.

This perspective also changes the role of interaction design. The policy through which AI capability is delivered structures when humans reason, receive feedback, delegate, learn, and adapt. Where humans are expected to interpret, challenge, transfer, recover, or act without equivalent assistance, evaluation should therefore consider not only immediate human--AI performance but also the capabilities and dispositions carried forward. This does not require maximal independence from AI; it requires clarity about which human capacities remain functionally necessary and how alternative interaction trajectories affect them.

The broader challenge for Human--AI Interaction is therefore not only to build systems that help humans perform better now, but to understand the trajectories through which those systems shape the humans who will act next. Evaluating interactive AI is incomplete whenever it measures what AI enables without also asking what its use leaves humans able, disposed, and prepared to do afterwards.

\appendix
\section{Supplementary reanalysis results}
\label{app:supplementary}

This appendix reports additional robustness and sensitivity analyses for
Studies~3 and~4 that complement the results presented in the main text.
For Study~3, we provide implementation details for the small-sample
cluster-robust inference and examine the sensitivity of the unaided-exam
contrasts to alternative covariate and fixed-effect specifications. For
Study~4, we report alternative longitudinal specifications, the auxiliary
retention model, and the full delta-adjusted missing-data sensitivity analysis.

\subsection{Study 3: Small-sample inference and specification sensitivity}
\label{app:p3}

The primary Study~3 analyses preserve the original clustered regression specification and supplement the published CR1 inference with CR2 standard errors and Satterthwaite degrees of freedom \citep{pustejovsky2018}, together with a wild cluster restricted bootstrap \citep{cameron2008,roodman2019}.

CR2 inference was computed in R 4.3.3 using the \texttt{clubSandwich} package (version 0.5.10), clustering by classroom. Tutor--Base contrasts were estimated as linear combinations within the same fitted model and covariance matrix.

The wild cluster restricted bootstrap was computed using \texttt{fwildclusterboot} (version 0.14.3; \texttt{dqrng} version 0.3.2), with classroom as the clustering unit. The null hypothesis was imposed, Rademacher weights were drawn once per classroom, and the test statistic was the CR1 $t$-statistic using the WCR11 variant. The package's default small-sample adjustment multiplies the CR1 variance by
\[
\frac{G}{G-1}\frac{n-1}{n-k},
\]
which rescales the original and bootstrap statistics equally and therefore does not affect the bootstrap $p$-value. Two-sided symmetric $p$-values were based on $B=9{,}999$ bootstrap replications.

Tutor--Base contrasts were tested as the linear restriction
\[
H_0:\beta_T-\beta_B=0
\]
within the same fitted model. The random-number generators used by base R and \texttt{dqrng} were reset with seed 20260928 before each of the six contrasts so that each bootstrap $p$-value is reproducible independently of estimation order.

\paragraph{Sensitivity to covariate and fixed-effect specification.}
The small-sample analyses reported in the main text show that the unaided-exam contrasts are sensitive to the inferential treatment of the 44 classroom clusters. We additionally examine whether the point estimates themselves depend on the covariate and fixed-effect specification. Table~\ref{tab:app-p3-fe} reports the two focal exam contrasts under the published specification and after progressively removing teacher effects, grader effects, and other covariates. These analyses are supplementary to the CR2/Satterthwaite and wild-cluster inference reported in the main text and are intended to assess specification sensitivity rather than to define alternative primary models.

\begin{table}[h!]
\centering
\small
\caption{Sensitivity of the unaided-exam contrasts in Bastani et al.\ to
the covariate and fixed-effect specification (normalised grade scale,
0--1). CR1 standard errors are clustered by classroom as in the
published specification; inference here uses a $t_{43}$ reference
distribution rather than the normal reference implied by the original
table. Sample size is 2{,}848 unless indicated; it changes with the specification because rows with missing values on the included covariates are dropped.}
\label{tab:app-p3-fe}
\begin{tabular}{>{\raggedright\arraybackslash}p{3.3cm}cccccc}
\toprule
 & \multicolumn{3}{c}{GPT Base $-$ Control}
 & \multicolumn{3}{c}{GPT Tutor $-$ GPT Base} \\
\cmidrule(lr){2-4}\cmidrule(lr){5-7}
Specification & Est. & 95\% CI & $p$ & Est. & 95\% CI & $p$ \\
\midrule
Published specification
& $-0.054$ & {\scriptsize $[-0.100,\,-0.009]$} & $0.020$
& $0.050$ & {\scriptsize $[0.003,\,0.097]$} & $0.038$ \\

Without teacher fixed effects
& $-0.020$ & {\scriptsize $[-0.062,\,0.022]$} & $0.348$
& $0.026$ & {\scriptsize $[-0.015,\,0.067]$} & $0.201$ \\

Without grader fixed effects ($n=2{,}849$)
& $-0.044$ & {\scriptsize $[-0.091,\,0.002]$} & $0.063$
& $0.041$ & {\scriptsize $[-0.003,\,0.086]$} & $0.065$ \\

Prior GPA, session and year only ($n=2{,}849$)
& $-0.015$ & {\scriptsize $[-0.059,\,0.029]$} & $0.501$
& $0.015$ & {\scriptsize $[-0.025,\,0.055]$} & $0.443$ \\

No covariates ($n=2{,}899$)
& $-0.035$ & {\scriptsize $[-0.102,\,0.033]$} & $0.307$
& $0.028$ & {\scriptsize $[-0.042,\,0.099]$} & $0.424$ \\
\bottomrule
\end{tabular}
\end{table}

\FloatBarrier

\subsection{Study 4: Longitudinal robustness and missing-data sensitivity}
\label{app:p4}

The primary Study~4 analysis uses the MMRM described in the main text.
The following analyses assess the robustness of the horizon-specific and
temporal contrasts to alternative longitudinal specifications and to
departures from the missing-at-random assumption.

\paragraph{Alternative longitudinal specifications.}
Table~\ref{tab:app-p4-ci} compares the horizon-specific residual
contrasts across the primary MMRM, a random-intercept LMM, GEE,
inverse-probability-of-retention weighted GEE, and horizon-specific
ANCOVA models.

\begin{table*}[ht]
\centering
\small
\caption{Horizon-specific residual contrasts in Corbett and Tangen under alternative longitudinal specifications (point estimates with 95\% CIs). MMRM: primary model, $N=375$, unstructured covariance. LMM: random intercept, condition $\times$ horizon with centred baseline as a main effect only, Kenward--Roger df. GEE: exchangeable working correlation with robust standard errors. IPW-GEE: independence working correlation with inverse-probability-of-retention weights estimated from condition and pre- and post-intervention belief. ANCOVA: separate baseline-adjusted models on the participants available at each horizon ($n=375$, $356$, $325$).}
\label{tab:app-p4-ci}
\begin{tabular}{lccccc}
\toprule
Contrast & MMRM & LMM & GEE & IPW-GEE & ANCOVA \\
\midrule
$D_{M-N}(\mathrm{post})$ & $-36.55$ & $-36.61$ & $-36.55$ & $-36.55$ & $-36.55$ \\
 & {\scriptsize $[-43.30,\,-29.80]$} & {\scriptsize $[-43.59,\,-29.63]$} & {\scriptsize $[-42.30,\,-30.80]$} & {\scriptsize $[-42.30,\,-30.80]$} & {\scriptsize $[-43.30,\,-29.80]$} \\
$D_{M-N}(\mathrm{10d})$ & $-29.63$ & $-29.52$ & $-29.59$ & $-29.51$ & $-29.02$ \\
 & {\scriptsize $[-36.90,\,-22.37]$} & {\scriptsize $[-36.63,\,-22.41]$} & {\scriptsize $[-36.40,\,-22.78]$} & {\scriptsize $[-36.45,\,-22.56]$} & {\scriptsize $[-36.42,\,-21.62]$} \\
$D_{M-N}(\mathrm{2m})$ & $-20.77$ & $-20.81$ & $-20.83$ & $-20.78$ & $-20.22$ \\
 & {\scriptsize $[-28.24,\,-13.31]$} & {\scriptsize $[-28.12,\,-13.50]$} & {\scriptsize $[-27.42,\,-14.25]$} & {\scriptsize $[-27.59,\,-13.97]$} & {\scriptsize $[-28.03,\,-12.41]$} \\
\addlinespace
$D_{M-T}(\mathrm{post})$ & $-9.77$ & $-9.73$ & $-9.77$ & $-9.77$ & $-9.77$ \\
 & {\scriptsize $[-16.51,\,-3.03]$} & {\scriptsize $[-16.70,\,-2.76]$} & {\scriptsize $[-17.72,\,-1.82]$} & {\scriptsize $[-17.72,\,-1.82]$} & {\scriptsize $[-16.51,\,-3.03]$} \\
$D_{M-T}(\mathrm{10d})$ & $-12.90$ & $-12.96$ & $-12.93$ & $-13.15$ & $-12.76$ \\
 & {\scriptsize $[-20.11,\,-5.69]$} & {\scriptsize $[-20.02,\,-5.91]$} & {\scriptsize $[-20.83,\,-5.02]$} & {\scriptsize $[-21.20,\,-5.11]$} & {\scriptsize $[-20.05,\,-5.46]$} \\
$D_{M-T}(\mathrm{2m})$ & $-8.21$ & $-8.20$ & $-8.20$ & $-7.47$ & $-7.53$ \\
 & {\scriptsize $[-15.53,\,-0.90]$} & {\scriptsize $[-15.36,\,-1.04]$} & {\scriptsize $[-16.19,\,-0.22]$} & {\scriptsize $[-15.83,\,0.89]$} & {\scriptsize $[-15.05,\,0.00]$} \\
\bottomrule
\end{tabular}
\end{table*}

\begin{table*}[ht]
\centering
\small
\caption{Delta-adjusted multiple-imputation sensitivity analysis for the two-month temporal contrasts in Corbett and Tangen (summary; all scenarios in Table~\ref{tab:app-p4-tip-full}). Imputed two-month values in the shifted arm were displaced by $\delta\in\{-30,-25,\ldots,30\}$ belief points; the other arms were imputed under missingness at random. The last column counts the values of $\delta$ for which the 95\% CI excludes zero. Equivalence was not re-evaluated under these scenarios.}
\label{tab:app-p4-tip}
\begin{tabular}{llcccc}
\toprule
Contrast & Shifted arm & Estimates & Min.\ lower CI & Max.\ upper CI & CI excl.\ 0 \\
\midrule
$\Delta D_{M-N}$ & Misconception AI & $12.03$ to $19.18$ & $4.92$ & $26.36$ & 13/13 \\
$\Delta D_{M-N}$ & Neutral AI & $9.56$ to $21.64$ & $2.33$ & $28.84$ & 13/13 \\
$\Delta D_{M-N}$ & Textbook & $15.58$ to $15.63$ & $8.44$ & $22.71$ & 13/13 \\
\addlinespace
$\Delta D_{M-T}$ & Misconception AI & $-2.16$ to $5.08$ & $-9.00$ & $11.99$ & 0/13 \\
$\Delta D_{M-T}$ & Neutral AI & $1.43$ to $1.49$ & $-5.50$ & $8.46$ & 0/13 \\
$\Delta D_{M-T}$ & Textbook & $-0.92$ to $3.84$ & $-7.79$ & $10.64$ & 0/13 \\
\bottomrule
\end{tabular}
\end{table*}

\paragraph{Delta-adjusted missing-data sensitivity.}
Table~\ref{tab:app-p4-tip} summarises the delta-adjusted
multiple-imputation analysis for the two primary two-month temporal
contrasts. Imputed two-month outcomes in one arm at a time are shifted
over the specified range of $\delta$ values, while the remaining arms
are imputed under missingness at random. The complete set of scenarios
is reported in Table~\ref{tab:app-p4-tip-full}.

\paragraph{Retention model and imputation procedure.}
The auxiliary retention model is a logistic regression of two-month
retention on condition (reference: Neutral AI), pre-intervention belief,
and post-intervention belief, fitted to all 375 participants
(Table~\ref{tab:app-p4-retention}). The delta-adjusted sensitivity
analysis used multiple imputation by chained equations (R package
\texttt{mice}) with Bayesian linear-regression imputation for the
10-day and two-month belief outcomes, with an imputation model that jointly included condition and the available pre-, post-, 10-day and two-month belief measures, each incomplete follow-up being imputed from the remaining variables, five iterations, and 40
imputed datasets (seed 20260928). Because this imputation model is unbounded, 17.1\% of the imputed two-month values and 15.7\% of the imputed 10-day values fell outside the 0--100 belief scale. Two bounded variants were therefore examined over the same grid of 39 scenarios: normal imputation and predictive mean matching (which imputes observed values), both with imputed values clamped to $[0,100]$ after the shift. Both reproduced the pattern of the primary analysis: every 95\% CI for $\Delta D_{M-N}$ excluded zero (estimates $12.44$--$21.42$) and every 95\% CI for $\Delta D_{M-T}$ included it (estimates $-1.53$ to $4.42$). Under these bounded variants, the share of the $M$--$N$ attenuation corresponding to movement of the Neutral arm ranged from 53\% to 64\% across Neutral-arm shifts from $0$ to $+30$ points, compared with 39\% to 62\% under the unbounded primary imputation.

After imputation, the imputed two-month values of the shifted arm only
were displaced by $\delta$. In each completed dataset, the temporal
contrasts were estimated from a change-score model in which the
two-month minus post-intervention belief score was regressed on
condition and pre-intervention belief; estimates were then combined
using Rubin's rules. At $\delta=0$, this model gives
$\Delta D_{M-N}=15.60$ and $\Delta D_{M-T}=1.46$, close to the primary
MMRM estimates.

\begin{table}[ht]
\centering
\small
\caption{Logistic regression of two-month retention in Corbett and Tangen ($N=375$; 325 retained). Odds ratios with Wald 95\% CIs; belief effects per 10 points on the 0--100 scale. Reference category: Neutral AI Dialogue.}
\label{tab:app-p4-retention}
\begin{tabular}{lccc}
\toprule
Predictor & Odds ratio & 95\% CI & $p$ \\
\midrule
Textbook Refutation (vs.\ Neutral AI) & 3.93 & [1.60, 9.64] & .003 \\
Misconception AI (vs.\ Neutral AI) & 2.65 & [1.11, 6.36] & .029 \\
Pre-intervention belief (+10 points) & 0.80 & [0.58, 1.11] & .186 \\
Post-intervention belief (+10 points) & 1.09 & [0.96, 1.23] & .169 \\
\bottomrule
\end{tabular}
\end{table}

\paragraph{Full delta-adjusted sensitivity results.}
Table~\ref{tab:app-p4-tip-full} reports every delta-adjusted scenario
used in the sensitivity analysis. The analysis is intended to assess
the robustness of the primary temporal contrasts to systematic
departures from missing at random; equivalence was not re-evaluated
under these scenarios.

\begin{table}[ht]
\centering
\footnotesize
\caption{All scenarios of the delta-adjusted sensitivity analysis in Corbett and Tangen: estimates and 95\% CIs (Rubin's rules, 40 imputations) of the two-month temporal contrasts when the imputed two-month values of one arm are shifted by $\delta$ belief points.}
\label{tab:app-p4-tip-full}
\begin{tabular}{lcc}
\toprule
$\delta$ & $\Delta D_{M-N}(\mathrm{post},2m)$ & $\Delta D_{M-T}(\mathrm{post},2m)$ \\
\midrule
\multicolumn{3}{l}{\textit{Shifted arm: Misconception AI}} \\
$-30$ & $12.03$ $[4.92, 19.14]$ & $-2.16$ $[-9.00, 4.68]$ \\
$-25$ & $12.62$ $[5.55, 19.70]$ & $-1.56$ $[-8.36, 5.25]$ \\
$-20$ & $13.22$ $[6.18, 20.26]$ & $-0.95$ $[-7.72, 5.82]$ \\
$-15$ & $13.82$ $[6.79, 20.84]$ & $-0.35$ $[-7.10, 6.40]$ \\
$-10$ & $14.41$ $[7.40, 21.42]$ & $0.25$ $[-6.48, 6.99]$ \\
$-5$ & $15.01$ $[8.00, 22.01]$ & $0.86$ $[-5.87, 7.58]$ \\
$0$ & $15.60$ $[8.60, 22.61]$ & $1.46$ $[-5.27, 8.19]$ \\
$+5$ & $16.20$ $[9.19, 23.21]$ & $2.06$ $[-4.68, 8.80]$ \\
$+10$ & $16.80$ $[9.76, 23.83]$ & $2.67$ $[-4.09, 9.42]$ \\
$+15$ & $17.39$ $[10.33, 24.45]$ & $3.27$ $[-3.51, 10.05]$ \\
$+20$ & $17.99$ $[10.90, 25.08]$ & $3.87$ $[-2.95, 10.69]$ \\
$+25$ & $18.58$ $[11.45, 25.71]$ & $4.48$ $[-2.39, 11.34]$ \\
$+30$ & $19.18$ $[12.00, 26.36]$ & $5.08$ $[-1.83, 11.99]$ \\
\addlinespace
\multicolumn{3}{l}{\textit{Shifted arm: Neutral AI}} \\
$-30$ & $21.64$ $[14.44, 28.84]$ & $1.43$ $[-5.50, 8.36]$ \\
$-25$ & $20.64$ $[13.50, 27.78]$ & $1.43$ $[-5.43, 8.30]$ \\
$-20$ & $19.63$ $[12.54, 26.72]$ & $1.44$ $[-5.38, 8.26]$ \\
$-15$ & $18.62$ $[11.57, 25.67]$ & $1.44$ $[-5.33, 8.22]$ \\
$-10$ & $17.62$ $[10.59, 24.64]$ & $1.45$ $[-5.30, 8.20]$ \\
$-5$ & $16.61$ $[9.60, 23.62]$ & $1.45$ $[-5.28, 8.19]$ \\
$0$ & $15.60$ $[8.60, 22.61]$ & $1.46$ $[-5.27, 8.19]$ \\
$+5$ & $14.60$ $[7.58, 21.61]$ & $1.46$ $[-5.27, 8.20]$ \\
$+10$ & $13.59$ $[6.56, 20.62]$ & $1.47$ $[-5.29, 8.23]$ \\
$+15$ & $12.58$ $[5.52, 19.65]$ & $1.48$ $[-5.32, 8.27]$ \\
$+20$ & $11.58$ $[4.47, 18.69]$ & $1.48$ $[-5.36, 8.32]$ \\
$+25$ & $10.57$ $[3.40, 17.74]$ & $1.49$ $[-5.41, 8.38]$ \\
$+30$ & $9.56$ $[2.33, 16.80]$ & $1.49$ $[-5.47, 8.46]$ \\
\addlinespace
\multicolumn{3}{l}{\textit{Shifted arm: Textbook}} \\
$-30$ & $15.63$ $[8.56, 22.70]$ & $3.84$ $[-2.96, 10.64]$ \\
$-25$ & $15.63$ $[8.58, 22.67]$ & $3.44$ $[-3.33, 10.21]$ \\
$-20$ & $15.62$ $[8.60, 22.65]$ & $3.05$ $[-3.70, 9.80]$ \\
$-15$ & $15.62$ $[8.60, 22.63]$ & $2.65$ $[-4.09, 9.39]$ \\
$-10$ & $15.61$ $[8.61, 22.62]$ & $2.25$ $[-4.48, 8.98]$ \\
$-5$ & $15.61$ $[8.61, 22.61]$ & $1.86$ $[-4.87, 8.58]$ \\
$0$ & $15.60$ $[8.60, 22.61]$ & $1.46$ $[-5.27, 8.19]$ \\
$+5$ & $15.60$ $[8.59, 22.61]$ & $1.06$ $[-5.67, 7.80]$ \\
$+10$ & $15.59$ $[8.57, 22.62]$ & $0.67$ $[-6.09, 7.42]$ \\
$+15$ & $15.59$ $[8.54, 22.64]$ & $0.27$ $[-6.50, 7.04]$ \\
$+20$ & $15.59$ $[8.51, 22.66]$ & $-0.13$ $[-6.92, 6.67]$ \\
$+25$ & $15.58$ $[8.48, 22.68]$ & $-0.52$ $[-7.35, 6.31]$ \\
$+30$ & $15.58$ $[8.44, 22.71]$ & $-0.92$ $[-7.79, 5.95]$ \\
\bottomrule
\end{tabular}
\end{table}

\FloatBarrier

\section*{Data and code availability}
The original datasets are publicly available from the repositories cited for each study \citep{vicente2025data,cabitza2026data,bastani2025data,corbett2026data}. Code that reconstructs the analytic datasets from these sources and reproduces all secondary analyses reported for Studies~2--4 (Study~1 uses only published aggregate results), together with the expected numerical output, is available at \url{https://github.com/ari-dasci/S-ResidualHumanStates}. For Study~4, the acquisition script removes direct identifiers (including IP addresses, geolocation and Prolific IDs) from the downloaded Qualtrics exports before analysis; Studies~2 and~3 use the analysis files released by the original authors. No participant-level data are redistributed.

\section*{Declaration of generative AI and AI-assisted technologies in the manuscript preparation process}

During the preparation of this work the authors used Claude (Anthropic) to assist with writing and verifying the R and Python code for the secondary reanalyses, cross-checking reported numbers against the analysis outputs, and revising the wording and structure of the manuscript. After using this tool, the authors reviewed and edited the content as needed and take full responsibility for the content of the published article.

\bibliographystyle{plainnat}
\bibliography{Residual_Human_States}
\end{document}